\documentclass[runningheads]{svjour2}
\smartqed  
\journalname{} 
\usepackage{amsmath, amsfonts, amsthm, amssymb} 
\input epsf 
\numberwithin{equation}{section}  
\usepackage{mathptmx} 
\newcommand \Mcal {\mathcal M} 
\newcommand \brho {\underline \rho} 
\newcommand \bsigma {\underline \sigma} 

\newcommand \bl {\underline l}
\newcommand \nb {\overline n}
\newcommand \bC {\underline C}
\newcommand \Cb {\overline C}

\newcommand \wt {\widetilde w} 
\newcommand \Ft {\widetilde F} 
\newcommand \bM {\underline M} 
\newcommand \bT {\underline T} 
\newcommand \Tb {\overline T} 
\newcommand \Mb {\overline M} 
\newcommand \barB {\underline B} 
\newcommand \Bbar {\overline B} 
\newcommand \barA {\underline A} 
\newcommand \Abar {\overline A}

\newcommand \bc {\underline c} 
\newcommand \bU {\underline U} 
\newcommand \bPhi {\underline \Phi} 
\newcommand \bPsi {\underline \Psi} 
\newcommand \cb {\overline c} 
\newcommand \ba {\underline a} 
\newcommand \bpsi {\underline \psi} 
\newcommand \psib {\overline \psi} 
\newcommand \bNcal {\underline {\mathcal N}} 
\newcommand \Ncalb {\overline {\mathcal N}} 

\newcommand \del        \partial
\newcommand \RR         {\mathbb{R}}

\newcommand \Pcal       {\mathcal P}  
\newcommand \Bcal       {\mathcal B}  
\newcommand \Ncal       {\mathcal N}

\newcommand \eps        \epsilon 
\newcommand \lam        \lambda   
\renewcommand \geq \geqslant
\renewcommand \leq \leqslant
\newcommand \Dcal 	{\mathcal D}

\newcommand \be   {\begin{equation}} 
\newcommand \ee   {\end{equation}}

\newcommand \osigma  {{\overline\sigma}} 
\newcommand \oepsilon  {{\overline\epsilon}} 
\newcommand \om  {{\overline m}}
\newcommand \oP  {{\overline P}}
\newcommand \odelta  {{\overline \delta}}

\newcommand \oalpha  {{\overline \alpha}}

\newcommand \olambda  {{\overline \lambda}}
\newcommand \tildepi  {{\widetilde \pi}}
\newcommand \tildemu  {{\widetilde \mu}}
\newcommand \tildelambda  {{\widetilde \lambda}}
\newcommand \tildealpha  {{\widetilde \alpha}}
\newcommand \tildebeta  {{\widetilde \beta}}
\newcommand \tildeoverpi  {{\widetilde {\overline\pi}}}
\newcommand \tildeoverbeta  {{\widetilde{\overline \beta}}}
\newcommand \tildeoveralpha {{\widetilde{\overline \alpha}}}
\newcommand \half    {\tfrac{1}{2}}
\newcommand \quarter    {\tfrac{1}{4}}
\newcommand \otop   {\stackrel{\circ}}  
\newcommand \Psib    {\overline \Psi}
\newcommand \Phib    {\overline \Phi}

\newcommand \ab    {\overline a}
\newcommand \bbar    {\overline b}
\newcommand \barb    {\underline b}
\newcommand \Ub    {\overline U}
\newcommand \Vb    {\overline V}

\let\oldmarginpar\marginpar
\renewcommand\marginpar[1]{\-\oldmarginpar[\raggedleft\footnotesize #1]%
{\raggedright\footnotesize #1}}

\begin{document}

\title{The characteristic initial value problem for 
plane symmetric spacetimes with weak regularity  
}
\titlerunning{The characteristic problem for plane symmetric spacetimes with weak regularity}

\author{Philippe G. LeFloch      \and
        John M. Stewart 
}

\institute{Philippe G. LeFloch \at
            Laboratoire Jacques-Louis Lions 
\& Centre National de la Recherche Scientifique, 
Universit\'e Pierre et Marie Curie (Paris 6), 4 Place Jussieu, 75252 Paris, France.   \\
              \email{pgLeFloch@gmail.com}         
           \and
         John M. Stewart \at
         Department of Applied Mathematics and Theoretical Physics, 
Centre for Mathematical Sciences, Cambridge CB3 0WA, UK.
 \\
              \email{J.M.Stewart@damtp.cam.ac.uk}
}

\date{}

\maketitle


\begin{abstract}
We investigate the existence and the global causal structure of plane sym\-metric spacetimes with 
weak regularity when the matter consists of an
  irrotational perfect fluid with pressure equal to its mass-energy density.  
Our theory encompasses the class of $W^{1,2}$ regular spacetimes 
whose metric coefficients have square-integrable first-order derivatives and 
whose curvature must be understood in the sense of distributions. 
We formulate the characteristic initial value problem with data posed on two
null hypersurfaces intersecting along a two-plane.
Relying on Newman-Penrose's formalism and expressing our weak regularity conditions in terms of the 
Newman-Penrose scalars, we arrive at a fully geometrical formulation in which, 
along each initial hypersurface,  
two scalar fields describing the incoming radiation must be prescribed in $L^1$ and $W^{-1,2}$, respectively.
To analyze the future boundary of such a spacetime and identify its global causal structure, 
we introduce a gauge that reduces the Einstein equations to a coupled system of wave equations and 
ordinary differential equations
for well-chosen unknowns. We prove that, within the {\sl weak regularity} class 
under consideration and for {\sl generic} initial data, a true spacetime singularity forms in finite proper time. 
Our formulation is robust enough so that propagating discontinuities in the 
curvature or in the matter variables do not prevent us from constructing a spacetime whose 
curvature generically blows-up on the future boundary.   
  Earlier work on the problem studied here  
  was restricted to sufficiently regular and vacuum spacetimes. 
\end{abstract}
  

\section{Introduction}
\label{INTRO} 

Our objective in this paper is to establish the existence of a large class of spacetimes with {\sl weak regularity}, 
and to shed some light on the propagation and interaction of gravitational waves. 
Specifically, we consider plane symmetric spacetimes satisfying the Einstein field equations of
general relativity
when the matter consists of an irrotational perfect fluid, whose pressure equals its mass-energy density.  
This latter assumption implies that the sound speed coincides with the light speed
 and, therefore,  discontinuities in the fluid variables may
occur only along null hypersurfaces.

Our aim is to construct a large class of plane symmetric spacetimes 
by formulating the characteristic initial value problem when data (with weak regularity)  
are prescribed on two null hypersurfaces intersecting along a two-plane. 
In addition, we investigate the regularity and global causal structure of such spacetimes, 
and establish the strong version of Penrose's cosmic censorship conjecture for 
spacetimes with {\sl weak regularity.}
Earlier work on this problem was restricted to sufficiently regular and vacuum spacetimes.  
As far as the matter model under consideration is concerned, 
the initial value problem was tackled earlier, especially by Taub \cite{Taub} (see also \cite{TT})  
who considered the initial value problem on a comoving hypersurface
and derived an integral formula for sufficiently regular solutions.

We are primarily concerned with the characteristic initial value problem and 
its well-posedness when, on each of the two hypersurfaces,
two scalar fields representing the gravitational radiation and the mass-energy density 
of the fluid are prescribed.  
Our presentation, based in part on the Newman-Penrose (NP) formalism, yields a geometrical  
formulation for the characteristic initial value problem 
in which physically relevant (i.e.~chart-invariant)
quantities are clearly identified. 
With regard to earlier related works, we want to mention that the characteristic initial value problem, 
addressed in the present paper for matter spacetimes with symmetry and low regularity,  
was treated earlier within the class of sufficiently regular vacuum spacetimes
 by 
Friedrich~\cite{Friedrich}, Stewart and Friedrich \cite{SF},  Rendall~\cite{Rendall}, and 
Christodoulou \cite{Christodoulou5}. 

We now outline what is meant by ``weak regularity'' in this paper. 
First of all, we emphasize that throughout Sections 2.1 to 3.4 we assume the spacetime to be regular. In Section 3.1 we show that  spacetimes of the symmetry class studied here are determined by the solution of singular wave equations
(cf.~\eqref{foo}, below), 
 for which the concept of weak solution is well-defined.   
We shall call spacetimes generated by these weak solutions ``weakly regular''. 
Equations of the form \eqref{foo}
have been studied extensively and are know as Euler-Poisson-Darboux equations.  We shall write down an integral representation of solutions of a certain characteristic initial value problem in terms of the Riemann-Green function, which is known explicitly, and the initial data appropriate to this problem, thus giving a representation of solutions both strong and weak solutions.

Most importantly, our results assume weak regularity on the data and,
therefore, on the solutions, so that the Einstein equations must be expressed in
the sense of distributions. Specifically, we introduce the {\sl class of $W^{1,2}$ regular spacetimes}
by requiring that the metric coefficients (in a certain gauge) 
belong to the Sobolev space $W^{1,2}$ of functions with square-integrable first-order derivatives.  
The low regularity imposed on the metric still is sufficient
for the Einstein equations to make sense within the theory of distributions \cite{LM}.
The Riemann curvature tensor then belongs to the dual space $W^{-1,2}$
 consisting of distributions that are derivatives of square-integrable functions.   
 
More precisely, in view of the Einstein equations, the Ricci part of the curvature turns out to be
solely in $L^1$ (that is, integrable) 
and may contain propagating jump discontinuities (along null hypersurfaces of the spacetime), 
while the Weyl part is solely a distribution in the negative Sobolev
space $W^{-1,2}$ and, for instance, may contain Dirac masses supported on null hypersurfaces. 
For further results on the existence and qualitative 
properties of weakly regular spacetimes, we refer to 
LeFloch and Rendall \cite{LR}, LeFloch and Smulevici \cite{LeFlochSmulevici}, 
and LeFloch and Stewart \cite{LS} (see also \cite{BLSS}).

The importance of spacetimes with very low regularity was first recognized by Christodoulou, 
who constructed (cf.~\cite{Christo00}--\cite{Christo2}) 
spherically symmetric spacetimes with matter, 
provided a detailed description of their global structure, 
and established the strong version of the cosmic censorship conjecture (within the restricted 
class of spacetimes under consideration).  
In comparison, the plane symmetry assumption made in the present paper
 leads to a technically simpler set of partial differential equations 
but yet allows us to study gravitational waves (and their interactions) which do not arise in spherical symmetry.

To investigate the global structure of the spacetimes under consideration, we introduce a gauge in which 
the reduced Einstein equations take the form of a system of 
wave equations with singular coefficients coupled to ordinary differential equations. 
The key equation is of the Euler-Poisson-Darboux type which we
solve with the Riemann function technique. The novelty of our analysis lies in the weak regularity assumed
by the solutions, as the initial data solely belong to the space $W^{1,2}$ and  
solutions must be understood in the distribution sense.  Uniqueness holds in the 
symmetry and weak regularity class under consideration. 

Within the low regularity class under consideration, we 
construct a development of a given initial data set
and we study the possible formation of curvature singularities in this spacetime
as one approaches its future boundary:  
we prove that {\sl generic} characteristic data always lead to a curvature singularity. 
To this aim, 
we study the blow-up behavior of the Riemann function and derive the leading 
behavior of certain curvature scalars,
which generically are found to, indeed, blow-up to infinity near the future boundary of the constructed spacetimes. For sufficiently regular and vacuum spacetimes, a related result was established earlier 
by Moncrief \cite{Moncrief} for Gowdy spacetimes with torus topology. 
Typical examples of this behavior (in the vacuum) were also constructed earlier explicitly 
by Khan and Penrose~\cite{KhanPenrose} and Aichelburg and Sexl \cite{AS}. 

Interestingly, the spacetimes constructed in the present work
can be interpreted as ``colliding spacetimes" 
in which two plane gravitational waves propagating in a flat
Minkowski background collide. Such spacetimes have been constructed explicitly for special choices of 
initial data sets \cite{AB,CX}, while other authors investigated general stability properties of colliding spacetimes  
\cite{KS,Penrose,Penrose0,Szekeres}, even going beyond the plane-symmetric case \cite{Yurtsever1,Yurtsever2}. 
Furthermore, the characteristic initial value problem was treated in \cite{Griffiths,GSR} when the initial data set is sufficiently regular. More recently, the fomation of trapped surfaces was analyzed in \cite{Feinstein,Ferrari,VW}. 
All of these earlier works were concerned with regular initial data sets, and 
our results in this paper generalize some of these results to the broad class of weakly regular spacetimes.

An outline of the structure of this paper follows.  In Section 2 we describe our matter model and write down the Einstein field equations for a regular spacetime.  In the first part of Section 3 we set up a characteristic initial value problem where the evolution equations are two uncoupled linear second order hyperbolic equations, and we obtain an explicit integral representation of their solutions in terms of appropriate initial data.  There is a well-defined concept of weak solutions for these evolution equations and this leads in the second half of Section 3 to our definition of $W^{1,2}$ space-times, and Theorem1, the existence theorem.  These results are local and depend on a particular choice of chart, and so in Section 4 we recast the results in chart-independent form using the Newman-Penrose (NP) notation.
In Section 5 we obtain our global existence theorem and investigate the details of curvature blowup.  In the final section we investigate jump conditions across a null hypersurface where the metric fails to be regular and study the propagation of curvature singularities.


\section{Einstein equations for plane symmetric spacetimes}
\label{F-0}

\subsection{The Einstein field equations} 
\label{21} 

In this section we first present our assumptions and write down Einstein
field equations for the geometrical set-up and matter model under consideration; we mostly follow
the presentation in Tabensky and Taub~\cite{TT}.  
Throughout Sections 2.1 to 3.4, the metric and fluid variables are assumed to be regular solutions of the
 field equations we are deriving.  Starting in Section 3.5 we consider weak solutions 
(defined in the sense of distributions). 
We are interested in plane symmetric spacetimes $(\Mcal, g)$
---or polarized Gowdy spacetimes---
described by  
\be 
  \begin{aligned} 
   g 
    & = e^{2a} \, (dt^2 - dx^2) - e^{2b} \, (e^{2c} \, dy^2 + e^{-2c} \, dz^2)
    \\
    & = e^{2a} \, du dv - e^{2b} \, (e^{2c} \, dy^2 + e^{-2c} \, dz^2), 
  \end{aligned}
  \label{F-metric}
\ee
where the scalars $a,b,c$ depend only upon the characteristic
variables $u= t-x$, $v=t+x$. Throughout this paper
{\sl we use the signature} $(+, -, -, -)$, which is standard in the physics literature. 
Observe that, while the functions $a$ and $c$ are coordinate-dependent, the function $b$ carries 
a geometric-meaning and
 $e^{b}$ represents the area element of the surfaces of symmetry
described by the coordinates $(y,z)$ and, if the coordinates on the orbits of symmetry 
are rescaled so that $(y,z) \mapsto (k y, kz)$ for some $k>0$, then 
$e^{b} \mapsto e^b/k$. 
 
The Einstein tensor $G_{\alpha\beta}$ associated with the metric \eqref{F-metric}, together with 
the energy-momentum tensor of the fluid $T_{\alpha\beta}$,
must satisfy the field equations  
\be
  G_{\alpha\beta} = \kappa \, T_{\alpha\beta}, \qquad \quad \alpha, \beta =0, \ldots, 3, 
  \label{F-Einstein}
\ee
where $\kappa>0$ is a constant.  
It is straightforward (but tedious) to compute the Christoffel symbols
and curvature coefficients  
associated with the metric \eqref{F-metric}, and to arrive at the
following expression of the relevant components of the Einstein tensor:  
\be
  \begin{aligned} 
    G_{00} & = 2 \, \bigl( 2 \, a_u b_u - b_{uu} - b_u^2 - c_u^2 \bigr), 
    \\
    G_{01} & = 2 \, \bigl( b_{uv} + 2 \, b_u b_v \bigr), 
    \\
    G_{11} & = 2 \, \bigl( 2 \, a_v b_v - b_{vv} - b_v^2 - c_v^2 \bigr), 
    \\
    G_{22} & = - 4 e^{-2a+2b+2c} \, \bigl( a_{uv} + b_{uv} + b_u b_v - b_u
    c_v - b_v c_u - c_{uv} + c_u c_v \bigr),  
    \\ 
    G_{33} & = - 4 e^{-2a+2b-2c} \, \bigl( a_{uv} + b_{uv} + b_u b_v + b_u
    c_v + b_v c_u + c_{uv} + c_u c_v \bigr), 
  \end{aligned} 
  \label{F-EinsteinTensor}
\ee 
where the subscripts on $a$, $b$ and $c$ represent partial derivatives
with respect to $u,v$. 
Recall that indices are lowered or raised using the metric, for
instance $G_{\alpha\beta} := g_{\alpha\gamma} g_{\beta\delta} G^{\gamma\delta}$.  

We assume that the matter is {\sl irrotational null fluid,} 
that is, an irrotational  
perfect fluid whose pressure $p$ is equal to its mass-energy density $w$, i.e. 
\be
  p= w. 
  \label{F-equationofstate}
\ee
It is described by the energy-momentum tensor 
(with our choice of signature)   
\be 
  T^{\alpha\beta} = (w +p) \, u^\alpha u^\beta - p \, g^{\alpha\beta}, 
  \label{F-energy}
\ee 
where $u^\alpha$ denotes the $4$-velocity vector of the fluid, normalized so that 
$$
  u^\alpha u_\alpha = 1.
$$ 
As we will see below, the condition $p=w$ implies
that the sound speed in the fluid coincides with the light speed, normalized to be~$1$.  

The (contracted) Bianchi identities for the Einstein tensor,
$\nabla_\alpha G^{\alpha\beta} = 0$,  
in combination with the field equations \eqref{F-Einstein} are equivalent to
the Euler equations  
$$
  \nabla_\alpha T^{\alpha\beta} = 0.
$$
Under our symmetry assumptions and for the matter model under consideration, the Euler
equations are equivalent to the  following partial differential equations
\be
  2 w_{,\alpha} \, u^\alpha u^\beta + 2w \, {u^\alpha}_{; \alpha} u^\beta 
  + 2w \, u^\alpha {u^\beta}_{;\alpha} - w_{,\alpha} \, g^{\alpha\beta} = 0, 
  \label{F-pde} 
\ee
where, for instance in ${u^\beta}_{;\alpha}$, the subscript ${}_{;\alpha}$ denotes the covariant derivative.

On the one hand, we can multiply \eqref{F-pde} by $u_\beta$ and obtain
the scalar equation  
$$
 2 w_{,\alpha} \, u^\alpha - 2w \, {u^\alpha}_{; \alpha} 
+ 2w \, u^\alpha u_\beta {u^\beta}_{;\alpha} - w_{,\alpha} \, u^\alpha = 0, 
$$
which, in view of the identity $u_\beta {u^\beta}_{;\alpha} = 0$,
simplifies into   
$$
u^\alpha w_{,\alpha} + 2w \, {u^\alpha}_{; \alpha} = 0. 
$$
Assuming that the density is bounded away from 
zero and setting 
$$
\Sigma = \half  \log w, 
$$
we conclude that  
\be
u^\alpha \Sigma_{,\alpha} + {u^\alpha}_{; \alpha} = 0. 
\label{F-scalar}
\ee

On the other hand, we can multiply \eqref{F-pde} by the projection
operator $H_{\beta\gamma} = g_{\beta\gamma} - u_\beta u_\gamma$   
and obtain the vector-valued equation 
$$
  H^{\alpha\gamma} w_{,\alpha} - 2w \, u^\alpha {u^\gamma}_{;\alpha} = 0, 
$$
or equivalently 
\be 
  H^{\alpha\gamma} \Sigma_{,\alpha} - u^\alpha {u^\gamma}_{;\alpha} = 0. 
  \label{F-vector}
\ee
 
In addition, we assume the flow to be irrotational and we introduce a
potential $\psi$ associated with the velocity, whose gradient is
timelike $\psi_{,\beta}\psi^{,\beta} > 0$: 
\be
  u_\alpha = \frac{\psi_{,\alpha}}{\sqrt{\psi_{,\beta} \, \psi^{,\beta}}}. 
  \label{C-upsi}
\ee 
The fluid equations \eqref{F-scalar} and \eqref{F-vector} then become
\be
  \left( \frac{w^{1/2} \, \psi^{,\alpha}}{
      \sqrt{\psi_{,\beta} \,\psi^{,\beta}}} \right)_{; \alpha} = 0 
  \label{F-scalar2}
\ee
and 
\be 
  \left( \Sigma - \log \sqrt{\psi^{,\alpha} \psi_{,\alpha}}\right)_{,\beta} 
  = k \, \psi_{,\beta},  
  \label{F-vector2}
\ee
respectively, 
where the scalar $k$ is 
$$
k = {\psi^{,\alpha} \Sigma_{,\alpha} \over \psi^{,\alpha} \psi_{,\alpha}}
    - {\psi^{,\alpha} \psi^{,\beta} \psi_{;\alpha\beta} \over
      (\psi^{,\alpha} \psi_{,\alpha})^2}. 
$$

The second equation, \eqref{F-vector2}, states that 
the gradient of ${\Sigma - \log \sqrt{\psi^{,\alpha}\psi_{,\alpha}}}$
is parallel to the gradient of $\psi$, so that 
the former can be expressed as $F(\psi)$ for some function $F$.  
By replacing $\psi$ by some function $G(\psi)$ if necessary we can always arrange that 
$\Sigma - \log \sqrt{\psi^{,\alpha} \psi_{,\alpha}} = 0$, in other words
\be
  w = \psi^{,\alpha} \psi_{,\alpha}. 
  \label{F-Bernoulli}
\ee 
This is the relativistic analogue of Bernoulli's law for irrotational flows in
classical fluid mechanics.  
It determines the mass-energy density algebraically, once we know the
velocity of the fluid.  

Finally, the equation \eqref{F-scalar2} determines the evolution of the
remaining fluid variable, that is,  
the potential $\psi$. 
By using the short-hand notation $\psi_{\alpha} = \psi_{,\alpha}$, $\psi_u= \psi_{,u}$,
etc.,  
and in view of 
$$
  \psi_{\alpha} = (\psi_u, \psi_v, 0,0), 
  \qquad 
  \psi^{\alpha} = 2 e^{-2a} \, (\psi_v, \psi_u, 0,0),
$$ 
it follows that  
$$
  \begin{aligned}
    \psi^{\alpha} \, \psi_{\alpha} & = 4 e^{-2a} \, \psi_u \psi_v, 
    \\
    u^\alpha  = {e^{-a} \over \sqrt{\psi_u \psi_v}} \, (\psi_v, \psi_u, 0,0), 
    & \qquad 
    u_\alpha  =  {e^{a} \over 2 \sqrt{\psi_u \psi_v}} \, (\psi_u,
    \psi_v, 0,0).    
  \end{aligned} 
$$ 
Hence, the equation \eqref{F-Bernoulli} becomes 
\be
\label{F-Bernoulli3}
  w = 4 e^{-2a} \, \psi_u \psi_v,  
\ee 
while \eqref{F-scalar2} reduces to a wave equation\footnote{This is a general fact for null fluids, irrespective of our symmetry assumption.} for the potential: 
\be
  \Box \psi = \psi_{uv} + b_v \, \psi_u + b_u \psi_v = 0. 
  \label{F-psi}
\ee 
The latter is the essential matter equation to be  investigated. 

Two main assumptions were used in our derivation: we needed that $w$ remains positive, and that 
the solutions are sufficiently regular. 
We will show how to relax the first of these in Section~2.2.  The regularity of solutions will be discussed in 
Section~3.5 and subsequent sections. 

We are now in a position to write down Einstein's field equations for the geometry variables. 
The components of the tensor 
$T_{\alpha\beta} = w \, \bigl( 2 \, u_\alpha u_\beta - g_{\alpha\beta}\bigr)$
are 
$$
  \begin{aligned} 
    & T_{00} = 2 \psi_u^2, \qquad T_{01} = 0, \qquad T_{11} = 2 \psi_v^2, 
    \\
    & T_{22} = 4 e^{-2a+2b+2c} \, \psi_u \psi_v,
    \qquad 
    T_{33} = 4 e^{-2a+2b-2c} \, \psi_u \psi_v.
  \end{aligned}  
$$
Returning to \eqref{F-Einstein} and relying on the expressions 
\eqref{F-EinsteinTensor} of the Einstein tensor, we arrive at the (evolution and constraint) 
equations for the metric coefficients $a,b,c$~: 
\be
  \begin{aligned}
    2 \, a_u b_u - b_{uu} - b_u^2 - c_u^2 & = \kappa \, \psi_u^2, 
    \\
    2 \, a_v b_v - b_{vv} - b_v^2 - c_v^2 & = \kappa \, \psi_v^2, 
    \\
    b_{uv} + 2 \, b_u b_v & = 0,
    \\ 
    - a_{uv} - b_{uv} - b_u b_v + b_u c_v + b_v c_u + c_{uv} - c_u c_v & = 
    \kappa \, \psi_u \psi_v,\\ 
    a_{uv} + b_{uv} + b_u b_v + b_u c_v + b_v c_u + c_{uv} + c_u c_v  & = 
    - \kappa \, \psi_u \psi_v.  
  \end{aligned} 
  \label{F-Einstein1}
\ee
Observe that the first three equations contain second-order derivatives of $b$, while 
the last two equations are equivalent to the system 
\be
 c_{uv} + b_u c_v + b_v c_u = 0, 
\label{F-Einstein2}
\ee 
\be
a_{uv} - b_u b_v + c_u c_v  = - \kappa \, \psi_u \psi_v,
\label{F-Einstein2b}
\ee
which contain second-order derivatives of $c$ and $a$, respectively. 
At this stage of the analysis, we observe that 
all of the equations under consideration are {\sl nonlinear}
 and involve quadratic products of first-order derivatives of the metric coefficients. 
This completes the derivation of the Einstein equations in characteristic coordinates
for
plane symmetric spacetimes $(\Mcal, g)$.  


\subsection{Physical meaning of the matter model}
\label{399}
 
The reader will have noticed that our derivation above assumed that $w$ defined in \eqref{F-Bernoulli}
remains positive. This condition may be imposed initially on the given data but, in general, will fail
after some finite time. When this happens,  
the four-velocity is no longer well-defined by \eqref{C-upsi}, and the
energy density $w$  given by \eqref{F-Bernoulli} becomes zero or negative.
In order to interpret the solution, one must return to the Euler equations \eqref{F-pde}
and realize that, when $w<0$, it is not possible
to normalize the velocity vector,   
but we can still express the {\sl 
energy-momentum tensor \eqref{F-energy} directly in terms of the potential} $\psi$, that is, 
\be
  T_{\alpha\beta} = 2 \, \psi_{\alpha} \psi_{\beta} 
  - (\psi^{\gamma} \psi_{\gamma}) \, g_{\alpha\beta}. 
  \label{F-energy3}
\ee 
Importantly, this expression is well-defined and {\sl regular} for all values of $\psi^\alpha \psi_\alpha$.
Moreover, if $\psi_v= \psi_{,v}$ vanishes so that $w$ vanishes, then the only non-vanishing component $T_{\alpha\beta}$ 
is  
$$
  T_{uu} = 2 \, (\psi_u)^2,
$$
which coincides with the energy-momentum tensor of  so-called null dust matter. 

The regime $w<0$ is most easily understood in terms of comoving coordinates.
Suppose first that $w=\psi^{\alpha}\psi_{\alpha}>0$ so that $\psi^{,\alpha}$
is a timelike vector (due to our choice $(+,-,-,-)$ for the signature). 
The comoving coordinates $(T,X,Y,Z)$ are defined as follows. 
Set $T=\psi(u, v)$, $Y=y$, $Z=z$, and define the function $X=X(u,v)$ via
$$
  dX = e^{b(u,v)}(\psi_{v}\,dv - \psi_{u}\,du).
$$
The integrability condition for such a solution to exist is precisely \eqref{F-psi}.
Noting that $dT = \psi_{u}\,du + \psi_{v}\,dv$, we obtain the spacetime metric 
in comoving coordinates
\be
  \label{eq:comov2} 
    g 
    = \frac{e^{2a}}{\psi^{\alpha}\psi_{\alpha}}
  \left(dT^{2} - e^{-2b}dX^{2}\right) - 
    e^{2b}\left(e^{2c}dY^{2}+ e^{-2c}dZ^{2}\right).
\ee
When $\psi^{\alpha}\psi_{\alpha}>0$, the variables $T,X$ are timelike 
and spacelike coordinates, respectively. 

Within the above setting, we can now consider the regime ${\psi^{\alpha}\psi_{\alpha}<0}$, 
for which $T$ is now a spacelike
coordinate and $X$ a timelike one.
Let $e^T_\alpha, e^X_\alpha, e^Y_\alpha$ and
$e^Z_\alpha$ be the corresponding unit vector fields so that
$$
    g = e^X_\alpha \, e^X_\beta - e^T_\alpha \, e^T_\beta 
    - e^Y_\alpha \, e^Y_\beta - e^Z_\alpha \, e^Z_\beta.
$$
Then, the energy-momentum tensor \eqref{F-energy3} takes the form 
$$
  T_{\alpha\beta} = (-\psi^{\gamma}\psi_{\gamma}) \, \big(e^X_\alpha \, e^X_\beta + e^T_\alpha \, e^T_\beta 
    - e^Y_\alpha \, e^Y_\beta - e^Z_\alpha \, e^Z_\beta \big), 
$$ 
which corresponds to a matter with a positive energy
density $\wt : = -\psi^\gamma \psi_\gamma > 0$ and an {\sl anisotropic} stress  
tensor with eigenvalues $\wt$, $-\wt$, and $-\wt$. This tensor does satisfy the weak energy condition and, 
therefore, should be regarded as ``physical''.


The above property can also be established by observing that, in the regime under consideration above, the fluid equations
reduce to the one of a scalar field. We refer the reader to Christodoulou~\cite{Christo2} for a detailled
 discussion of equations of state for fluids. 
From now on, we regard $\psi$ as the main fluid unknown, and \eqref{F-energy3} as 
the main expression of the energy-momentum tensor. 


\section{The characteristic problem for metrics with weak regularity}
\label{F-00} 

\subsection{Normalization and choice of coordinates} 

We can take advantage of the coordinate freedom to simplify radically
the set of nonlinear equations \eqref{F-psi} and \eqref{F-Einstein1} derived in the previous section. 
Namely, as we show now, the metric coefficient $c$ and the velocity potential 
$\psi$ can be regarded as the essential variables and are governed by singular wave equations.  
 
Solving the third equation in \eqref{F-Einstein1}, that is, $b_{uv} + 2 \, b_u b_v = 0$,
is straightforward, since it is equivalent to the wave equation 
${(e^{2b})_{uv} =0}$.  
Hence, there must exist functions $f,g$ such that 
$$
  e^{2b} = f(u) + g(v) >0. 
$$
Since the transformations $u \mapsto U(u)$ and $v \mapsto V(v)$ do not change the form of the metric \eqref{F-metric},  
we may choose the coordinates $u,v$ to coincide with the functions $f$ and $g$, respectively.
For definiteness, we consider the case that both $f$ and $g$ are {\sl decreasing,} so that $b$ decreases toward the future
and a singularity is expected in finite time in the future direction.  
It is convenient then to adopt the normalization 
\be
  f(u) = - \half \, u, \qquad g(v) = - \half \, v, 
\ee
in order to easily recover certain particular solutions available in the literature. 
Then, the metric coefficient $b$ is simply 
\be 
  b = \half  \log \big( \half \, |u+v|\big), 
  \label{F-functionb2}
\ee  
and the set $\bigl\{ u+v < 0 \bigr\}$ is the region of physical
interest, while the hypersurface $u+v=0$  corresponds to a (physical or
coordinate) singularity.  Hence, we are treating here the situation that $b$ is decreasing toward the future.

With this choice of coordinates, the spacetime $(\Mcal, g)$ is described by 
the following remaining equations:  
\be
  \begin{aligned} 
    & 2 (u+v) \, \psi_{uv} + \psi_u + \psi_v = 0, 
    \\
    & 2 (u+v) \, c_{uv} + c_u + c_v = 0, 
    \\
    & a_u = (c_u^2 + \kappa \,  \psi_u^2) \, (u+v) - \tfrac{1}{4} (u+v)^{-1},
    \\
    & a_v = (c_v^2 + \kappa \, \psi_v^2) \, (u+v) - \tfrac{1}{4}(u +
    v)^{-1},
    \\ 
    & a_{uv} = - c_u c_v  - \kappa \, \psi_u \psi_v + \tfrac{1}{4} (u+v)^{-2}. 
  \end{aligned} 
  \label{F-keyset} 
\ee 
Thus, each of the functions $\psi$ and $c$ satisfies the same singular 
wave equation, which is a special case of the Euler-Poisson-Darboux
(EPD) equations 
\be
\label{foo}
  \psi_{uv} + \frac{\alpha\psi_{u}}{u+v} +
  \frac{\beta\psi_{v}}{u+v} + \frac{\gamma\psi}{(u+v)^{2}}=0,
\ee
with given constants $\alpha, \beta, \gamma$. Such equations were first discussed 
systematically by Darboux \cite{Darboux} (Chap.~III and IV); 
for more recent material, see \cite{ALF,Stewart09}.  

Once $c$ and $\psi$ are determined by the first two equations in \eqref{F-keyset}, 
one determines the coefficient $a$ from the third and fourth
equations. Note that the compatibility condition
$(a_u)_v = (a_v)_u$ is then automatically satisfied by virtue of the  
first two equations in \eqref{F-keyset}. 
The last equation for $a_{uv}$ in \eqref{F-keyset} is redundant.

Once $a,c,\psi$ are determined, the mass-energy density $w$ is recovered by
Ber\-noul\-li's law \eqref{F-Bernoulli3}, and when $w>0$ the fluid velocity is given by \eqref{C-upsi}.  


\subsection{The essential field equation in a characteristic rectangle}
\label{secEPD}

We now begin our analysis of the first equation (for instance) 
in \eqref{F-keyset}, i.e.  
\be
  \label{eq:crg1}
  L[\psi]:= \psi_{uv}+\half(u+v)^{-1}(\psi_{u}+\psi_{v}) = 0,
\ee 
which (with some abuse of notation) we refer to as the essential field equation. 
It is appropriate to pose the characteristic initial value problem.
Given ${u_0<u}$ and ${v_0<v}$, let $P$ be the two-plane $(u,v)$ and $S$ be the two-plane
$(u_0,v_0)$. The plane $P$ is assumed to lie in the chronological future of $S$, that is, 
$u >u_0$ and $v >v_0$. 

Based on the past of $P$ and the future of $S$, it is natural to introduce 
the two-planes $R=(u_0, v)$ and $Q=(u,v_0)$ and the associated region 
$$
\Dcal = \Dcal(u_0, v_0; u,v) \subset \Mcal 
$$
defined as the diamond-shaped region with boundary $PQSRP$.  
The value of $\psi$ is then specified on each of the initial hypersurfaces 
$$
\bNcal = \bNcal(u_0, v_0; u,v) := SR, \qquad \qquad \Ncalb = \Ncalb(u_0, v_0; u,v) := SQ,
$$ 
and we aim at deriving an explicit representation of $\psi(P)
=\psi(u,v)$ in terms of these data.
Since the coefficients of \eqref{eq:crg1} are singular on the hypersurface 
$u+v=0$, we assume that $u_0+v_0<u+v<0$, that is
 $\Dcal$ lies to the past of this hypersurface.
Later, we will examine the behavior of the solutions as $u+v\to 0-$.
%

%
%
%

\begin{remark} 
\label{rem31} 
1. Throughout this paper, one could assume $0<u_0+v_0<u+v$ with virtually no change in the forthcoming analysis, 
and this would indeed be relevant for an analysis in the past direction. 

2. It is worth keeping in mind that not all solutions of \eqref{eq:crg1} become singular on the line $u+v=0$, and 
an obvious counter-example is $\psi(u,v) = 3u^{2}-2uv + 3v^{2}$.  
However,  $\psi$ does become singular for generic initial data, as we will show later 
(Section~\ref{54section}). 
\end{remark}


\subsection{Representation formula} 

Our analysis of the equation \eqref{eq:crg1} is based on the Riemann function approach. 
From now on, we regard the coordinates $(u,v)$ of $P$ as
fixed and we introduce a new independent variable $(u',v') \in \Dcal$.
The {\sl operator adjoint to $L$} (defined in \eqref{eq:crg1})
applies to functions $\varphi=\varphi(u',v')$ 
and reads 
\be
  \label{eq:crg2}
  L^*[\varphi] := \varphi_{u'v'}-\half(u'+v')^{-1}(\varphi_{u'}+\varphi_{v'}) +
  (u'+v')^{-2}\varphi,
\ee
which is also of Euler-Poisson-Darboux type. 
By setting  
\be
  \label{eq:crg3}
\theta(u',v') := {\varphi(u',v') \over u'+v'},
\ee
the adjoint equation $L^*[\varphi]=0$ is found to be equivalent to 
\be
  \label{eq:crg4}
  \theta_{u'v'} + \half(u'+v')^{-1}(\theta_{u'}+\theta_{v'})=0, 
\ee
which {\sl coincides} with the original operator \eqref{eq:crg1}.

We are going to construct a special solution of $L^*[\varphi]=0$
which satisfies a backward characteristic initial value problem,
now with data posed on $PQ$ and $PR$. Specifically,  we choose these data to be  
\be
  \label{eq:crg5}
  \varphi(u,v')=\left(\frac{u+v'}{u+v}\right)^{1/2}\qquad
  \text{ on } PQ, \quad v_0\leqslant v' \leqslant v,
\ee
and 
\be
  \label{eq:crg6}
  \varphi(u',v)=\left(\frac{u'+v}{u+v}\right)^{1/2}\qquad
  \text{ on } PR, \quad u_0\leqslant u' \leqslant u.
\ee
The reason for this choice will become clear shortly. 
This solution depends on the independent variable $(u',v')$, 
as well as on the fixed parameter $(u,v)$, and it is convenient to   
write $\varphi=\varphi(u',v';u,v)$ to indicate this dependence.
The solution $\varphi$ of \eqref{eq:crg3}-\eqref{eq:crg4} satisfying \eqref{eq:crg5}-\eqref{eq:crg6} is commonly called the \emph{Riemann function}.
Such a solution $\varphi$ does exist since \eqref{eq:crg4} is a {\sl
  linear} partial differential equation with regular coefficients
  since the diamond $\Dcal$, by assumption, does not intersect
the singularity.  

In view of the definitions of $L$ and $L^*$, we have
$$
\aligned
  0 & = 2 \, (\varphi L[\psi] - \psi L^*[\varphi])
   \\
   &=  \Big(
  (\varphi\psi)_{u'} - 2 \psi\varphi_{u'}+(u'+v')^{-1}\psi\varphi\Big)_{v'}
   + 
           \Big(
           (\varphi\psi)_{v'} - 2 \psi\varphi_{v'}+(u'+v')^{-1}\psi\varphi\Big)_{u'}, 
\endaligned
$$ 
which we now integrate over $\Dcal$ by using Stokes'
theorem to convert the right-hand side to line integrals.
After a straightforward integration by parts and dividing by a factor $2$, we obtain 
$$
\aligned
    & (\psi\varphi)(P) - (\psi\varphi)(Q)
    - (\psi\varphi)(R) + (\psi\varphi)(S)
    \\
    &+ \left(\int_{Q}^{P} + \int_{R}^{S}\right)
    \psi \, \Big( \half (u'+v')^{-1}\varphi - \varphi_{v'}\Big) \, dv'
    \\
    &+ \left(\int_{R}^{P} + \int_{Q}^{S}\right)
    \psi \, \Big( \half (u'+v')^{-1}\varphi - \varphi_{u'}\Big) \, du' = 0.
\endaligned
$$
In view of the initial data \eqref{eq:crg5} assumed by $\varphi$, we
see that the 
contribution from the segment $QP$ vanishes, while similarly 
\eqref{eq:crg6} implies that the integral over $RP$ vanishes. 
We also use that these two pieces of initial data for $\varphi$ imply $\varphi(P)=1$.

Thus, we find the following formula for the general solution of \eqref{eq:crg1} 
\be
\label{eq:crg999}
\aligned 
\psi(u,v) 
= 
    \, &  \varphi(u,v_0;u, v)\psi(u, v_0) + 
    \varphi(u_0,v;u, v)\psi(u_0, v)
   \\
   &  - \varphi(u_0,v_0;u, v)\psi(u_0, v_0) 
     - \int_{u_0}^{u} \psi(u',v_0) \, \barA[\varphi](u', v_0; u,v) \, du'
    \\
    & - \int_{v_0}^{v} \psi(u_0,v') \, \Abar[\varphi](u_0, v'; u,v] \, dv',
  \endaligned
\ee
which is the promised representation in terms of data on $\bNcal$ and $\Ncalb$, 
with 
\be
\label{eq:crg10-bis00}
\aligned
\barA[\varphi](u', v_0; u,v)  :=& \varphi_{u'}(u',v_0; u,v) - \half (u'+v_0)^{-1}\varphi(u',v_0; u,v), 
\\
\Abar[\varphi](u_0, v'; u,v] :=& \varphi_{v'}(u_0,v'; u,v) - \half (u_0+v')^{-1}\varphi(u_0,v'; u,v). 
\endaligned
\ee  

Sufficient regularity is assumed for the time being. Later on,   
from the regularity of the Riemann function it will follow that the above formula 
makes sense as long as the data $\psi(u_0, \cdot)$ and $\psi(\cdot, v_0)$ are locally integrable.
Furthermore, provided 
the initial data are more regular and admit locally integrable, first-order derivatives, 
an integration by parts gives an alternative (and somewhat simpler) representation: 
\be
  \label{eq:crg10}
  \aligned
    \psi(u,v) = \,   \varphi(u_0,v_0;u, v) \, \psi(u_0, v_0) 
    &+ \int_{u_0}^{u}\varphi(u',v_0;u,v) \, \barB[\psi](u', v_0) \, du'
     \\
    &+ \int_{v_0}^{v}\varphi(u_0,v';u,v) \, \Bbar[\psi](u_0, v'] \, dv',
  \endaligned
\ee
where $\psi(u_0, v_0)$ and 
\be
\label{eq:crg10-bis}
\aligned
\barB[\psi](u', v_0)  :=& \psi_{u'}(u',v_0) + \half (u'+v_0)^{-1}\psi(u',v_0), 
\\
\Bbar[\psi](u_0, v'] :=& \psi_{v'}(u_0,v') + \half (u_0+v')^{-1}\psi(u_0,v'), 
\endaligned
\ee 
are given by the 
prescribed characteristic data.


\subsection{Riemann function for the essential field equation}

The formula \eqref{eq:crg10} would have no utility, unless we can construct 
the Riemann function explicitly, which we now do. The results in this subsection are given in sufficient
detail that they can be verified by (tedious) calculations. The reasons \emph{why} the approach below works 
require further knowledge of the theory of the Euler-Poisson-Darboux equation, for which we refer to
\cite{Darboux}; see also \cite{ALF,Stewart09} and the references therein.

We start from the equation \eqref{eq:crg4} for the ``dual'' Riemann function $\theta$ and seek a 
homogeneous solution depending essentially upon the ratio $u'/v'$: 
\be
  \label{eq:crg11}
  \theta(u',v') = (v')^{-1/2}y(-u'/v').
\ee
Some elementary manipulation shows that $\theta$ satisfies
\eqref{eq:crg4} if and only if $y= y(z)$ satisfies
\be
  \label{eq:crg12}
  z(1-z)y''(z) + (1-2z)y'(z) - \tfrac{1}{4}y(z)=0. 
\ee
Interestingly, this is a particular case of the \emph{hypergeometric equation}, studied for instance 
in \cite{AS65} (Section~15.5) and \cite{Olver}. 

In order to impose the boundary conditions needed to identify the function $y$ we
need the following result (verifiable by brute force \footnote{Alternatively, one can observe that 
\eqref{eq:crg1} is the cylindrically symmetric wave equation in $(3+1)$ dimensions and that  \eqref{eq:crg13} is a conformal transformation and, in fact, an inversion.}): if
$\theta(u',v')$ satisfies \eqref{eq:crg4}, then so does
\be
  \label{eq:crg13}
  {\widehat\theta}(u',v') = \frac{1}{(u'+v)^{1/2}(v'-u)^{1/2}}
  \, \theta\left(-\frac{u'-u}{u'+v}, \frac{v'+u}{v'-v}\right),
\ee
where, as before, $u$ and $v$ are regarded as parameters.
Thus, taking into account the transformations \eqref{eq:crg3}, \eqref{eq:crg11}, and \eqref{eq:crg13}
and recalling the boundary conditions \eqref{eq:crg5} and \eqref{eq:crg6}, we find 
\be
  \label{eq:crg14}
  \varphi(u',v'; u,v) = \frac{(u'+v')}{(u+v')^{1/2}(u'+v)^{1/2}} \, y(z),
\ee
where
\be
\label{zzz}
  z = z(u',v'; u,v) := \frac{(v'-v)(u'-u)}{(v'+u)(u'+v)}
\ee
and $y$ is a solution of \eqref{eq:crg12} satisfying  
$$
  y(0) = 1.
$$

Now the hypergeometric equation \eqref{eq:crg12}  has regular singular
points at $z=0$, $z=1$, and $z=\infty$. 
Near $z=0$, there exist two independent solutions with asymptotic forms 
$y_{1}(z)\sim z^{0}$ and $y_{2}(z)\sim z^{0}\log z$, respectively.
It is clear that the initial condition $y(0) = 1$ picks out unambiguously $y_{1}(z)$, i.e.,
\be
  \label{eq:crg17}
  y(z) = F(\half, \half; 1; z),
\ee
where, by definition, $F(a, b;c;z)$ is the standard \emph{hypergeometric function} 
and is regular for $|z| < 1$. 

Thus, we deduce from \eqref{eq:crg14} the following expression of the Riemann function: 
\be
  \label{eq:crg18}
  \varphi(u',v';u,v) = 
  \left(\frac{u'+v'}{u'+v}\right)^{1/2}
  \left(\frac{u'+v'}{u+v'}\right)^{1/2}
  F(\half, \half; 1; z),
\ee
Now on $PQ$ and $PR$ we clearly have $z=0$ and, in particular, $z(Q) = z(R)=0$.
Along $QS$, the variable $z$ increases monotonically to its maximum value $z(S)$.
Note that 
\be
  \label{eq:z}
  1 - z = \frac{(u'+v')(u+v)}{(u'+v)(v'+u)} \in [0,1), 
\ee
so that, as long as $u+v$ remains bounded away from zero, i.e., as long as 
$P$ does not approach the singular line $u+v=0$.
The condition $z\geq 0$ is obvious from \eqref{zzz}, while 
$z < 1$ is obvious from \eqref{eq:z}; 
then $z \in[0,1)$ remains bounded away from $1$, and
the Riemann function is {\sl regular}. This is true, in particular, on the lines $QS$ and $RS$. 

Inserting \eqref{eq:crg18} into \eqref{eq:crg10}, we arrive at the following main conclusion.

\begin{proposition}
\label{finform}
The solution of the characteristic initial value problem associated with the essential field
 equation \eqref{eq:crg1} is given by 
the general formula \eqref{eq:crg10}-\eqref{eq:crg10-bis} in terms of boundary data $\psi(\cdot, v_0)$
and $\psi(u_0, \cdot)$, in which the Riemann function reads  
\be
  \label{eq:crg18-ter}
  \varphi(u',v';u,v) = 
  \left(\frac{u'+v'}{u'+v}\right)^{1/2}
  \left(\frac{u'+v'}{u+v'}\right)^{1/2}
  F\left(\half, \half; 1; \frac{(v'-v)(u'-u)}{(v'+u)(u'+v)}\right)
\ee  
and the (hypergeometric) function $F$ is singular when its last argument tends to $1$. 
\end{proposition}


\subsection{Definition and existence of $W^{1,2}$ regular spacetimes}
\label{lowsec}

We now return to the full set of Einstein equations \eqref{F-keyset} or, more precisely, 
\be
  \begin{aligned} 
    & 2 (u+v) \, \psi_{uv} + \psi_u + \psi_v = 0, 
    \\
    & 2 (u+v) \, c_{uv} + c_u + c_v = 0, 
    \\
    & a_u = (c_u^2 + \kappa \,  \psi_u^2) \, (u+v) - \tfrac{1}{4} (u+v)^{-1},
    \\
    & a_v = (c_v^2 + \kappa \, \psi_v^2) \, (u+v) - \tfrac{1}{4}(u + v)^{-1},
  \end{aligned} 
  \label{F-keysetNEW} 
\ee
which we refer to as the {\sl reduced Einstein equations} for plane symmetric spacetimes 
$(\Mcal, g)$ (in the chosen gauge).  
Now, we are interested in encompassing solutions $\psi, c, a$ with {\sl weak regularity.}
Specifically, we propose to search for solutions $\psi, c$ in the Sobolev space $W^{1,2}$ of functions 
which are square-integrable together with their first-order derivatives. This class is natural since the curvature is then 
well-defined in the distributional sense, and the Einstein equations \eqref{F-Einstein} hold as equalities
between distributions in the dual space $W^{-1,2}$, as established in~\cite{LM}. 

Given any $u_0<u$ and $v_0<v$ with $u+v < 0$, we focus on the characteristic initial value problem
within the (regular) region $\Dcal=\Dcal(u_0, v_0; u,v) \subset \Mcal$ defined in Section~\ref{secEPD}.  
The boundaries $\bNcal$ and $\Ncalb$ are null hypersurfaces on which we prescribe data and which intersect
on the two-plane    
$$
 \Pcal = \Pcal(u_0, v_0) := \left\{ u=u_0, \, v=v_0 \right\}.  
$$ 
Without loss of generality, we can normalize the metric
coefficient $a$ to satisfy  
\be
  \label{aone}
  a(u_0,v_0) = 0 \qquad \text{ on the two-plane } \Pcal. 
\ee
At this stage, it is most convenient to provide a first description of
our results in terms of the (coordinate dependent) coefficients $a,c$ and potential $\psi$. 
But, later in Section~\ref{glob} (cf.~Theorem~\ref{F-global}) after some further analysis, 
we will restate these results in a {\sl fully geometric} (i.e., chart invariant) form.

It is convenient to introduce the following notation for any function $f=f(u,v)$ 
$$
\aligned 
\bM_{u_0, v_0}^{u,v}[f]
: =&  
\sup_{v_0 \leq v' \leq v} \Big( \int_{u_0}^u |f(\cdot, v')|^2 \, du'\Big)^{1/2},  
\\
\Mb_{u_0, v_0}^{u,v}[f]
: =& 
\sup_{u_0 \leq u' \leq u} \Big(\int_{v_0}^v |f(u',\cdot)|^2 \, dv'\Big)^{1/2}
\endaligned 
$$
and, by extension, 
$$
\bM_{u_0, v_0}^{u,v_0}[f] : = \Big( \int_{u_0}^u |f(u', v_0)|^2 \, du'\Big)^{1/2}, 
\qquad
\Mb_{u_0, v_0}^{u_0,v}[f] : = \Big(\int_{v_0}^v |f(u_0, v')|^2 \, dv'\Big)^{1/2}. 
$$  
Other straightforward extensions of this notation will be used, for instance
$$
\bM_{u_0, v_0}^{u,v}[f,h] = \bM_{u_0, v_0}^{u,v}[f] + \bM_{u_0, v_0}^{u,v}[h]
$$
if two functions $f, h$ are given.

\begin{definition}[Notion of $W^{1,2}$ regular spacetime] 
\label{Hone}
Given $u_0, v_0$ and $u, v$ with ${u<u_0}$, ${v<v_0}$ and ${u_0 + v_0 <0}$, 
a {\sl $W^{1,2}$ regular spacetime satisfying the Einstein
equations} \eqref{F-keysetNEW} in the characteristic rectangle $\Dcal=\Dcal(u_0, v_0; u,v)$ and 
describing self-gravi\-tating irrotational fluids, 
is determined in characteristic coordinates 
$(u,v)$ by three  
continuous metric coefficients $a,b,c$ (cf.~\eqref{F-metric})
and a continuous fluid potential $\psi$ (cf.~\eqref{C-upsi} and \eqref{F-Bernoulli}) 
such that:  

1. The coefficient $b$ is given by the explicit formula \eqref{F-functionb2}. 

2. The derivatives of $a, c, \psi$ are differentiable in a weak sense and the (semi-) norms  
$$
\bM_{u_0, v_0}^{u,v}[|a_u|^{1/2}, c_u, \psi_u], 
\qquad 
\Mb^{u_0, v_0}_{u,v}[|a_v|^{1/2}, c_v, \psi_v]  
$$
are finite.

3. The reduced field equations \eqref{F-keysetNEW} hold in the sense of distributions.

4. The function $a$ satisfies the normalization \eqref{aone}. 

\end{definition}

Our terminology ``$W^{1,2}$ regular spacetime'' is motivated by the fact that $c$ and $\psi$ are the essential unknowns
of the problem and, by our definition, have square-integrable first-order derivatives. 
As we observed earlier, initial data are required for 
the functions $c$ and $\psi$, only, while the function $a$ can be recovered afterwards.

Recall that the potential $\psi$ determines, both, the velocity field $u$
and the mass density $w$. In view of \eqref{F-Bernoulli3}, the regularity assumed in the above definition
implies that the spacetime integral of the mass density 
\be
\label{290}
\aligned
&  \iint_{\Dcal(u_0, v_0; u,v)} |w(u',v')| \, du'dv'
\\
&  \leq 4 \, e^{-2 \min_\Dcal a} \, (u_0 - u)^{1/2} \, (v_0 - v)^{1/2} \, 
  \bM_{u_0, v_0}^{u,v}[\psi_u] 
  \,  
   \Mb_{u_0, v_0}^{u,v}[\psi_v]   
\endaligned
\ee
is also finite. 
On the other hand, no general estimate is available for the velocity field,
since the norm of the gradient of $\psi$, that is, $w$, may well vanish, at which
point \eqref{C-upsi} is ill-defined. Recall that 
 $\psi$ is the main fluid variable, and
\eqref{F-energy3} is the primary expression of the energy-momentum tensor.

\begin{theorem}[Well-posedness theory]
\label{F-Cauchy}
Let $\Dcal=\Dcal(u_0, v_0;u,v)$ be a characteristic rectangle that does not intersect the singularity hypersurface.  
Let $\bpsi, \bc$ and $\psib, \cb$ be continuous functions (of a single
variable) defined on the null hypersurfaces  
$$
\bNcal = \left\{ u_0 < u' < u; \quad v'=v_0\right\}, 
\qquad 
\Ncalb
= \left\{u'=u_0; \quad v_0 < u' < u\right\},
$$
 respectively, 
and satisfying the continuity conditions $\bpsi(u_0) = \psib(v_0)$
and $\bc(u_0) = \cb(v_0)$, 
such that the semi-norms
$$
  \begin{aligned}
    &  \bM_{u_0, v_0}^{u,v_0}[\bc_u, \bpsi_u], 
  \quad 
    &  \Mb_{u_0, v_0}^{u_0,v}[\cb_v, \psib_v]   
  \end{aligned}
$$
are finite.  
Then, there exists a unique $W^{1,2}$ regular spacetime (in the sense of Definition~\ref{Hone}) 
determined by functions $a,b,c, \psi: \Dcal \to \RR$ that 
satisfy the reduced Einstein equations \eqref{F-keysetNEW} 
of self-gravitating irrotational fluids and assume the initial data 
$$
\aligned 
& c(\cdot,v_0) = \bc, \qquad \psi(\cdot, v_0) = \bpsi \qquad \text{ on the hypersurface } \bNcal, 
\\
& c(u_0, \cdot) = \cb, \qquad \psi(u_0, \cdot) = \psib \qquad \text{ on the hypersurface } \Ncalb.
\endaligned
$$
\end{theorem}

Some additional remarks about the regularity of our solutions are in order. Recall the embedding $W^{1,2} \subset C^{1/2}$ valid on one dimension, so that any solution is H\" older continuous 
in each characteristic variable.
 The second-order derivatives
of the solutions $\psi, c$ (which can always be defined in the sense of distributions) 
{\sl need not be functions} in a classical sense. However, the mixed derivatives $\psi_{uv}$ and $c_{uv}$ 
do have some regularity and, specifically, the spacetime integral
\be
\label{291}
  \iint_{\Dcal(u_0, v_0; u,v)} \big( |c_{uv}|^2 + |\psi_{uv}|^2 \big) \, du'dv'
\ee
is also finite. 

\begin{proof} Given the reduction analysis and 
the observations already made in the present section, the
  proof is now direct.  Indeed, we have derived the integral formula \eqref{eq:crg10}-\eqref{eq:crg10-bis},
 which provides an
explicit expression for the solutions $\psi, c$ to the first two equations in
\eqref{F-keysetNEW} in term of their characteristic data  $\bpsi,
\bc$ and $\psib, \cb$ prescribed on the null hypersurfaces $\bNcal$ and $\Ncalb$, respectively. 
Since the Riemann function is bounded (at least) in the domain $\Dcal$
under consideration, as explained before Proposition~\ref{finform},
all the integrals in \eqref{eq:crg10} make sense since $\barB[\psi]$ and 
$\Bbar[\psi]$ are (in $L^2$ and thus) integrable.  The existence of functions $\psi, c:
\Dcal \to \RR$ satisfying the reduced Einstein equations in the distribution sense
and the desired initial conditions is now clear. 

Furthermore, estimates on the norms of these solutions, as required in 
Definition~\ref{Hone}, can be established from the integral formulation, as follows.
Considering for instance the function $\psi$, for all $(u',v') \in \Dcal(u_0, v_0; u,v)$ we have 
$$
  \begin{aligned} 
\psi(u',v') 
 = \,   &  \varphi(u',v_0;u', v')\, \bpsi(u') 
    + 
    \varphi(u_0,v;u', v') \, \psib(v') 
    \\
    & - \varphi(u_0,v_0;u', v') \half  \left( \bpsi(u_0) + \psib(v_0) \right)   
     \\
    & - \int_{u_0}^{u'} \bpsi(u'') \, \barA[\varphi](u'',v_0;u',v') \, du'' 
    - \int_{v_0}^{v'} \psib(v'') \, \Abar[\varphi](u_0,v'';u',v') \, dv''. 
  \end{aligned}
$$
Using that the Riemann function is regular away from the singular hypersurface, 
we compute the derivative $\psi_{u'}$ and obtain 
$$
  \begin{aligned}
    | \psi_{u'}(u',v')|  \lesssim
    \, & \, |\bpsi_{u'}(u')| + |\bpsi(u')| + |\psib(v')| + |\bpsi(u_0)| + |\psib(v_0)|
     \\
    & + \int_{u_0}^{u'} |\bpsi(u'')| \, du'' + \int_{v_0}^{v'} |\psib(v'')| \, dv'', 
  \end{aligned}
$$
in which the notation $f \lesssim g$ means that there exists a
constant $C>0$ such that $f \leq C \, g$.  
Since a direct calculation allows us to control the sup norm of $\psi$ within the spacetime region, 
we
obtain a pointwise control  of the $u'$-derivative of $\psi$ in terms
of the same derivative of the initial data:  
$$
  | \psi_{u'}(u',v')|  \lesssim 
  \, |\bpsi_{u'}(u')| + \sup_{\Dcal(u_0, v_0; u,v)}  \big( |\bpsi| + |\psib| \big). 
$$
By integrating this inequality along an arbitrary characteristic line   
allows us to control the desired $W^{1,2}$ norm of the solution with the same norm of the initial data: 
$$
\aligned 
\bM_{u_0, v_0}^{u,v}[\psi_u]
=&  
\sup_{v_0 < v' < v} \Big( \int_{u_0}^u |\psi_u(\cdot, v')|^2 \, du'\Big)^{1/2}
\\
\leq & \, \bM_{u_0, v_0}^{u,v_0}[\bpsi_u] + \sup_{\Dcal(u_0, v_0; u,v)}  \big( |\bpsi| + |\psib| \big). 
\endaligned
$$
The same arguments apply to $\psi_v$, as well as to the coefficient $c$.

Next, we return to the key set of  equations \eqref{F-keysetNEW}  
and determine the metric function $a$ as follows.
By integrating the relevant equations in \eqref{F-keysetNEW}
along the initial hypersurfaces 
$\bNcal$ and $\Ncalb$, respectively, and using the normalization \eqref{aone} we obtain the initial 
values for the function $a$: 
$$
\ba(u'):=  a(u', v_0) = 
  \int_{u_0}^{u'} \left( 
    (\cb_{1,u}^2 + \kappa \, \bpsi_{u}^2)(u'') \, (u''+v_0) - \tfrac{1}{4}
    (u'' + v_0)^{-1} \right) \, du'',  
$$
$$
  \ab(v'):=a(u_0, v') = \int_{v_0}^{v'} \left( 
  (\cb_{2,v}^2 + \kappa \, \psib_{v}^2)(v'')  \, (u''+v'') - \tfrac{1}{4}
    (u'' + v'')^{-1} \right) \, dv''. 
$$
Both expressions provide bounded functions since the prescribed characteristic data belong to $W^{1,2}$. 
Having determined $a$ on the two null hypersurfaces $SR$ and $SQ$, we can  
use once more the last two equations in \eqref{F-keyset} and, by integrating along characteristics, 
determine the function $a$ within the whole spacetime region $\Dcal$. 
The equations \eqref{F-keysetNEW} show that the functions $a_u$ and $a_v$ have the same regularity as 
the products $c_u \, \psi_u$ and 
$c_v \, \psi_v$, respectively, that is, are integrable along characteristic lines. Hence, 
$\bM_{u_0, v_0}^{u,v}[|a_u|^{1/2}]$ and $\Mb^{u_0, v_0}_{u,v}[|a_v|^{1/2}]$ are finite. 
\end{proof}


\section{The characteristic problem for NP scalars with weak regularity}
\label{C-0}

\subsection{Choice of the tetrad}
 
Our objective here is to reformulate more geometrically the characteristic initial value problem when data are
given on two intersecting null hypersurfaces and, specifically, to identify {\sl which physical data}
should be imposed on these hypersurfaces. 
To emphasize that the null coordinates under consideration {\sl can differ}
 from the ones constructed 
in the previous section, we denote them by $(U,V)$. 
We are going to re-derive the expressions of the field equations
(obtained in Section~\ref{F-0}) via the formalism introduced by
Newman and Penrose~\cite{NP}, following here the notation in \cite{Stewart91}. Our main conclusion in the present section, as stated in Proposition~\ref{theo-regul} below, is a restatement ---in the NP notation---
of the weak regularity conditions introduced in Definition~\ref{Hone}.

The tangent space can be described by a basis $n^\alpha$, $l^\alpha$,
$m^\alpha$, $\om^\alpha$ of {\sl complex-valued} null vectors satisfying the normalization: 
$$
\label{norm1} 
1= l^\alpha n_\alpha = n^\alpha l_\alpha = - m^\alpha \om_{\alpha} = - \om^\alpha m_\alpha, 
$$
while all other contractions vanish. 
In view of the expression \eqref{F-metric} of the metric for plane symmetric spacetimes, we choose the tetrad
$$
  l = \frac{\del}{\del U}, \quad   n = e^{-2a} \frac{\del}{\del V}, \quad
  m = \frac{e^{-b}}{\sqrt{2}} \,   \Big( e^{-c} \frac{\del}{\del y}
    +  i \, e^{c} \, \frac{\del}{\del z} \Big),  
$$
in which the treatment in the $U$- and $V$-directions is not symmetric
as this has some advantages in the applications (to conveniently prescribe the incoming radiation).  
As is standard, 
we use the notation $D, \Delta, \delta, \overline \delta$ for the directional derivatives
associated with the null tetrad.  

We begin by describing general formulas satisfied by  Newman-Penrose scalars in such a null frame and, next, 
will specify our particular choice of tetrad. 

\subsection{Expressions for the Newman-Penrose scalars}

In the NP notation, the Christoffel symbols associated with the metric
\eqref{F-metric} are represented by the following twelve {\sl connection NP scalars:}  
\be
  \begin{aligned}
    & \alpha = \beta = \gamma = 0, \quad &&\eps = a_U, 
    \\
    & \kappa =0, 
    \quad &&\lam =e^{2a}\, c_V, \\
    & \mu = e^{2a} \, b_V, \quad  &&\nu = \pi =0, 
    \\
    & \rho = - b_U, \quad &&\sigma = - c_U, \quad &&&\tau = 0,
  \end{aligned} 
\ee 
while the Ricci curvature is represented by the following seven {\sl Ricci curvature NP scalars}:
\be 
  \begin{aligned}
    & \Phi_{00} = e^{-4a} \, \bigl( b_{UU} - 2 \, 
    a_U b_U + b_U^2 + c_U^2 \bigr), 
    \\
    & \Phi_{01} = \Phi_{12} = 0, 
    \\ 
    & \Phi_{02} = e^{-2a} \, \bigl( c_{UV} + b_U c_V + b_V c_U \bigr), 
    \\
    &\Phi_{11} = \half   e^{-2a} \, 
    \bigl( - a_{UV} + b_U b_V - c _U c_V \bigr), 
    \\ 
    & \Phi_{22} =  e^{4a} \, \bigl (- b_{VV} + 2 \, 
    a_V b_V - b_V^2 - c_V^2 \bigr ), 
    \\
    & \Lambda = \tfrac{1}{6} e^{-2a} \, 
    \bigl( a_{UV} + 2 b_{UV} +3 b_U b_V + c_U c_V \bigr),
  \end{aligned}
  \label{C-Phi}
\ee 
and the Weyl curvature (or free
radiation) is represented by the following {\sl five Weyl curvature NP scalars:} 
\be
  \begin{aligned}
    & \Psi_0 = - c_{UU} + 2 \, (a_U -  b_U) c_U, 
    \\
    & \Psi_1 = \Psi_3 = 0, 
    \\
    & \Psi_2 = \tfrac{1}{3} \,e^{-2a}\, \bigl( - a_{UV} + b_{UV} + 
    2 \, c_U c_V \bigr), 
    \\ 
    & \Psi_4 = e^{4a} \, \bigl(   - c_{VV} + 2 \, (a_V - b_V) c_V  \bigr). 
  \end{aligned}
  \label{C-Psi}
\ee 

Of course, the Ricci curvature is related to the matter tensor via the field equations  
\be
  \begin{aligned} 
    \Phi_{\alpha\beta} &= - \tfrac{\kappa}{2} \Big( T_{\alpha\beta} - 
    \tfrac{1}{4} T \, g_{\alpha\beta}\Big),  &&
    \\
    \Lambda &= - \tfrac{\kappa}{24} \, T, \qquad && T := T_\alpha^\alpha. 
  \end{aligned} 
  \label{C-Einstein5} 
\ee
To make this more explicit, it is convenient to decompose the fluid velocity vector $u^\alpha$ in the form 
\be
  u^\alpha = Z \, l^\alpha + W \, n^\alpha, \qquad 
  Z = u_\alpha n^\alpha, \quad 
  W = u_\alpha \, l^\alpha, 
\ee
so that the condition $u_\alpha \, u^\alpha = 1$ becomes $W Z = \half$. 
From the expression of the energy-momentum tensor (for general pressure $p$) 
$$
  T^{\alpha\beta} = (w +p) \, u^\alpha u^\beta - p \, g^{\alpha\beta}, 
  \qquad 
  T= w - 3 p, 
$$ 
and in view of \eqref{C-Einstein5} we can compute 
$$
  \begin{aligned}
    & \Phi_{\alpha\beta} = -\tfrac{\kappa}{2} (w + p) 
    \bigl( u_\alpha u_\beta - \tfrac{1}{4} \, g_{\alpha\beta} \bigr), 
    \\
    & \Lambda = - \tfrac{\kappa}{24} \, (w - 3 p). 
    \end{aligned}
$$
We thus find 
\be
  \Phi_{02} = 0, 
  \label{C-Phi02}
\ee
as well as the relations 
$$
  \begin{aligned}
  & \Phi_{00} = -\tfrac{\kappa}{2} \, (w + p) \, W^2, \quad 
  \Phi_{11} = - \tfrac{\kappa}{8} \, (w +p), 
  \quad 
  \Phi_{22} = -\tfrac{\kappa}{2} \, (w + p) \, Z^2,
  \\
      & \Lambda = - \tfrac{\kappa}{24} \, (w - 3 p), 
    \end{aligned}
$$ 
which, after imposing $p=w$ for a null fluid, become 
\be
\begin{aligned}
\Phi_{00} &= - \kappa \, w \, W^2, \quad 
  \Phi_{11} = - \tfrac{\kappa}{4} \, w, 
  \quad 
  \Phi_{22} = - \kappa \, w \, Z^2, 
  \\ 
  \Lambda & = \tfrac{\kappa}{12} \, w. 
\end{aligned}
  \label{C-Phi-b}
\ee 

In view of the third relation in \eqref{C-Phi}, the equation \eqref{C-Phi02} is equivalent to
a wave equation for the function $c$ 
$$
  c_{UV} + b_U c_V + b_V c_U = 0, 
$$
which allows us to recover the key equation \eqref{F-Einstein2} derived earlier.


\subsection{Evolution equations for the NP scalars} 

Before we can proceed further, we need to identify which variables are the essential dependent variables
that require initial data. 
First of all, it follows from \eqref{C-Phi-b} that $\Phi_{00}, \Phi_{22}, \Phi_{11}$ are not independent
but satisfy the constraint 
\be
  C_{1} := \big( \Phi_{00} \, \Phi_{22}\big)^{1/2} + 2 \, \Phi_{11} =0. 
  \label{C-phi22} 
\ee
(Recall that $\Phi_{00}$ and $\Phi_{22}$ are both real and have the same sign.) 
We thus impose the constraint $C_1 = 0$ everywhere and regard
$\Phi_{11}$ as a redundant dependent variable. 
For if we know $\Phi_{00}$ and $\Phi_{22}$ at a point 
where $w\geqslant 0$, then (4.8) determines $\Phi_{11}$ up to a sign, and (4.7) requires $\Phi_{11} \leqslant 0$.

Similarly, $\Phi_{11}$ and $\Lambda$ are not independent but are related
algebraically: 
\be
  C_{2} := \Phi_{11} + 3 \, \Lambda =0, 
  \label{C-Lambda} 
\ee
and thus $\Lambda$ can be viewed as a redundant dependent variable.
Interestingly, the relation \eqref{C-Lambda} is equivalent to an evolution equation for $b$, i.e.  
$$
  b_{UV} + 2 \, b_U \, b_V = 0,  
$$
which allows us to recover the equation \eqref{F-Einstein1}. 

Third, the NP scalars satisfy one further algebraic constraint because
$\delta\rho=\overline\delta\sigma=0$. By taking into account
the property $C_1 = C_2=0$ this constraint can be written in the form 
\be
  \label{eq:C-algebraic1}
  C_{3} := \Psi_{2} + 2 \Lambda - ( \rho\mu - \sigma\lambda) =0,
\ee
which means that $\Psi_{2}$ can also be regarded as a redundant
dependent variable.

We are now in a position to list the complete set of evolution equations for the following {\sl nine 
non-redundant NP scalars} 
$$
\aligned
& \epsilon, \, \rho, \, \sigma, \, \lambda, \, \mu, 
\\
& \Phi_{00}, \, \Phi_{22}, \, \Psi_0, \, \Psi_4. 
\endaligned
$$

First of all, the field equations consist of {\sl four evolution equations in
the $D$-direction:}
\be
  \begin{aligned}
    & D_{1} := D \rho - (\rho(\rho + 2\epsilon) + \sigma^2 + \Phi_{00}) =0, 
    \\
    & D_{2} := D \sigma - (2 \,(\rho + \epsilon) \sigma + \Psi_0)=0, 
    \\
    & D_{3} := D \lambda - ((\rho - 2\epsilon) \lambda + 
    \sigma \mu)=0, 
    \\
    & D_{4} := D \mu - 2(\rho - \epsilon) \mu=0,
  \end{aligned}
  \label{C-Dequations}
\ee
and {\sl five evolution equations in the $\Delta$-direction:} 
\be
  \begin{aligned}
    & \Delta_{0} := \Delta \epsilon - (- \rho\mu+\sigma\lambda + 6\Lambda) =0, 
    \\
    & \Delta_{1} := \Delta \rho + 2 \mu\rho =0,
    \\
    & \Delta_{2} := \Delta \sigma + \mu\sigma + \lambda \rho =0, 
    \\
    & \Delta_{3} := \Delta \lambda + 2\mu\lambda + \Psi_4 =0, 
    \\
    & \Delta_{4} := \Delta \mu + \mu^2 + \lambda^2 + \Phi_{22} =0. 
\end{aligned}
  \label{C-Deltaequations}
\ee
Note that the constraints $C_{1}=C_{2}=C_{3}=0$ have been imposed 
in the equations above, and will be imposed as well in the Bianchi
identities below. 
The relevant {\sl two contracted Bianchi identities} 
\be
  \label{Bianchi}
  \begin{aligned}
    & D_{5} := D \Phi_{22}  + 4\mu\Phi_{11} + 2(-\rho + 2\epsilon) \Phi_{22} =0, 
    \\
    & \Delta_{5} := \Delta\Phi_{00} + 2\mu\Phi_{00} - 4\rho\Phi_{11} =0, 
  \end{aligned}
\ee
while the {\sl two independent Bianchi identities} reduce to 
\be
  \label{C-Bianchi1}
  \begin{aligned}
    & D_{6} := D \Psi_4 - ( -3 \lambda \Psi_2 + (\rho - 4\epsilon)\Psi_4 - 
    2 \lambda \Phi_{11} + \sigma \Phi_{22}) =0,
    \\
    & \Delta_{6} := \Delta \Psi_{0} - ( -\mu\Psi_{0}+ 3\sigma\Psi_{2} - 
    \lambda\Phi_{00} + 2\sigma\Phi_{11}) =0.
  \end{aligned}
\ee

 It is important to notice that the quantities $D_{n}$ and $\Delta_{n}$ 
($n=1,\ldots, 4$) are not independent but satisfy {\sl four differential
identities.} To obtain these, observe that for any scalar $f$
$$
  D(\Delta f) - \Delta(D f) = -2 \epsilon \, \Delta f,
$$
showing that $D$ and $\Delta$ do not commute in general.
We can for instance express this general identity for the scalar $\rho$, 
and then rewrite $D\rho$
in terms of $D_{1}$ and $\Delta\rho$ in terms of $\Delta_{1}$.
Then we replace all $D$-derivatives of  NP quantities by the
appropriate $D_{k}$ and all $\Delta$-derivatives by the corresponding
$\Delta_{k}$. Finally, making use of the three constraint terms $C_{k}$ we arrive at 
the desired identity and, 
applying the same process to $\sigma$, $\lambda$ and $\mu$ in turn, we obtain
\be
\label{eq:allcom} 
\aligned 
    D \Delta_{1} - \Delta D_{1} 
& = 2\rho \Delta_{0} + 2\mu D_{1} + 2\rho \Delta_{1} +
    2\sigma \Delta_{2} + 2\rho D_{4} + \Delta_{5} + 4\rho C_{2},
\\ 
    D \Delta_{2} - \Delta D_{2} 
& = 2\sigma \Delta_{0} + \lambda D_{1} +2 \sigma
    \Delta_{1} + \mu D_{2} + 2\rho \Delta_{2}
     + \rho D_{3} + \sigma D_{4} 
     \\
    & \hskip4.5cm + 
    \Delta_{6} + 2\sigma C_{2} + 3\sigma C_{3},
\\ 
    D \Delta_{3} - \Delta D_{3} 
& = - 2\lambda \Delta_{0} + \lambda \Delta_{1} +
    + \mu \Delta_{2} + 2\mu D_{3} + (\rho-4\epsilon) \Delta_{3} 
     + 2\lambda D_{4} 
       \\
    & \hskip3.8cm + \sigma \Delta_{4} + 
    D_{6} - 2\lambda C_{2} - 3\lambda C_{3},
\\ 
    D \Delta_{4} - \Delta D_{4} 
& = - 2\mu \Delta_{0} + 2\mu \Delta_{1} +2 \lambda
    D_{3} + 2\mu D_{4} + 2\rho \Delta_{4} + D_{5} - 4\mu C_{2}.
\endaligned
\ee

Finally, once the NP scalars have been determined, the metric coefficients in \eqref{F-metric} are simply recovered by   
integrating out the following equations for $a,b,c$: 
\be
  \begin{aligned}
      & a_U = \epsilon,\\
      & b_U = - \rho, \qquad b_V = e^{-2a}\mu, \\ 
      & c_U = - \sigma, \qquad c_V = e^{-2a}\lambda.  
    \end{aligned}
    \label{C-metric}
\ee 
Observe that there is {\sl no equation for $a_V$.} (Recall that the coordinates $u$ and $v$ do not play symmetric role, 
due to our choice of tetrad.)


\subsection{Revisiting the characteristic initial value problem with regular data}

We are ready to formulate the characteristic initial value problem for plane symmetric spacetimes. 
We are given a point $(U_0, V_0)$ representing the two-plane    
$$
  \Pcal := \left\{ U=U_0, \, V=V_0 \right\}, 
$$  
and consider the two hypersurfaces $\bNcal = \left\{ V = V_0 \right\}$ and  
$\Ncalb = \left\{ U =U_0 \right\}$ with intersection $\Pcal$. 

The null coordinates and the tetrad are geometrically defined as follows. We first choose the vectors 
$\bl$ and $\nb$ at $\Pcal$ to be future-directed null vectors normalized by the condition $g(\bl, \nb) = 1$
and tangent to $\bNcal$ and $\Ncalb$, respectively. 
The vectors are determined up to the transformation $(\bl, \nb) \mapsto (p \bl, \nb/p)$ for 
$p>0$. 
The quotient manifold $Q$ of the four-dimensional 
spacetime by the symmetry group (i.e. the two-dimensional Euclidian group) is represented by each of the timelike surfaces
which are orthogonal to the group orbits. The vectors $\bl, \nb$ can then be extended as geodesic fields
in the future direction and, in the quotient picture,
 their integral curves are nothing but $\bNcal, \Ncalb$, respectively.
We then define the coordinate $U$ on $\bNcal$ to be the affine parameter of $\bl$ normalized so that 
$U=U_0$ at $\Pcal$, the coordinates $V$ is defined similarly from $\nb$. Next, the functions $U,V$ can be 
uniquely extended
to $Q$ by the requirement that their level sets are outgoing and incoming null curves, respectively. 

From the gradient of these functions, we can then determine $l^\alpha = g^{\alpha\beta} \del_\beta V$ and $n^\alpha = g^{\alpha\beta} \del_\beta U$, which are null geodesic fields satisfying the orthogonality conditions
$l^\alpha \del_\alpha V = 0= n^\alpha \del_\alpha U$. Hence, $l$ is tangential to $\bNcal$ and 
$n$ is tangential to $\Ncalb$. Consequently, along $\bNcal$ we can write $l = \bC \, \bl$ for some constant
$\bC>0$, while along $\Ncalb$ we have $n = \Cb \, \nb$ for some constant
$\Cb>0$. Then, we write $g(l,n) = g^{\alpha\beta} \del_\alpha V \del_\beta U= g^{UV} = e^{-2a}$, and 
we define $l' = e^{2a} l$ and $n' = e^{2a} n$. It follows that 
${l'}^\alpha \del_\alpha U = g(l',n) = e^{2a} g(l,n) = 1$ and, similarly, 
${n'}^\alpha \del_\alpha V = g(n',l) = e^{2a} g(n,l) = 1$. Consequently, we actually have 
$l' =\del_U$ and $n'=\del_V$, and it follows also that 
$l^\alpha \del_\alpha U= e^{-2a}$ and $n^\alpha \del_\alpha V = e^{-2a}$.

Now, along the hypersurface $\bNcal$ and in view of $l^\alpha \del_\alpha U= e^{-2a}$ and 
$\bl^\alpha \del_\alpha U=1$, we conclude that 
$\bC = e^{2a}$ along $\bNcal$.  
An analogous argument tells us also that $\Cb = e^{2a}$ $\Ncalb$ 
and, by continuity of the function $a$
at $\Pcal$, it follows that for some $k>0$ we have $\bC=\Cb=k$ and therefore 
$e^{-2a} = k$ along $\bNcal \cup \Ncalb$. Recalling that 
$l = \bC \, \bl$, $n = \Cb \, \nb$, and  $\bC=\Cb=k$, we get 
$\bl = (1/k) l$ and 
$\nb = (1/k) n$, so that 
$$
1 = g(\bl, \nb) = k^{-2} g(l,n) = k^{-2} e^{-2a} = 1/k. 
$$ 
 We conclude that $k=1$, which justifies the normalization that $a=0$ on $\bNcal \cup \Ncalb$.

Our formulation is based on prescribing the following NP scalars  
\be
\label{dataPcal}  
\rho, \sigma, \lambda, \mu  \quad \text{ on the plane } \Pcal,  
\ee 
together with data 
\be
  a, \Psi_0, \Phi_{00} \quad     \text{ on the hypersurface } \bNcal, 
  \label{C-dataN}
\ee
and 
\be
  a, \Psi_4, \Phi_{22}    \quad \text{ on the hypersurface } \Ncalb. 
  \label{C-dataNdash}
\ee   
 These data can be given freely, except that we require 
 $\Phi_{00} \leq 0$ on $\bNcal$  and $\Phi_{22} \leq 0$ on
 $\Ncalb$, in agreement with \eqref{C-Phi-b}. In addition, since regularity is required in the present 
 section, we impose that
 $$
 \ba(U_0 ) = \ab(V_0). 
 $$ 
Furthermore, without loss of generality we impose the normalization 
\be
\label{normaPcal}  
a=b=c=0 \quad \text{ on the plane } \Pcal. 
\ee 
Our aim is to determine the solution components
$$
\epsilon, \rho, \sigma,
\lambda, \mu,
\qquad 
 \Phi_{00}, \Phi_{22}, \Psi_{0},\Psi_{4}
$$
 in the future $\Dcal$ of $\bNcal\cup\Ncalb$. 

Consider first the solution on the two-plane $\Pcal$.
In view of \eqref{dataPcal} and \eqref{C-dataN} we know $\epsilon$, $\rho$, $\sigma$,
$\lambda$ and $\mu$.
Also since we know $\Phi_{00}$ and $\Phi_{22}$ at $\Pcal$,
\eqref{C-phi22} fixes $\Phi_{11} \leq 0$. 
The scalar $\Lambda$ follows from \eqref{C-Lambda}, and $\Psi_{2}$ from
\eqref{eq:C-algebraic1}. This determines the complete set of dependent variables on $\Pcal$.

Next consider the solution on the hypersurface $\bNcal$.
Since $a(\cdot, V_0)= \ba$ is prescribed, we know $\epsilon$ from \eqref{C-metric}.
Next, \eqref{C-Dequations} gives a coupled system of {\sl nonlinear} ordinary differential equations of Riccati
 type for $(\rho,\sigma)$, with known source terms and initial data known on the plane $\Pcal$. 
Hence, using the prescribed data at $(U_0, V_0)$ we see that this system
has a unique solution $(\rho, \sigma)$ defined on a maximal interval, denoted by $[U_0, \bU_0)$,
Returning to \eqref{C-Dequations}, we may then solve a {\sl linear} system for the unknowns $(\lambda,\mu)$ on $\bNcal$ with prescribed initial data on $\Pcal$.   
Next, using the condition $C_1=0$ to express $\Phi_{11} \leqslant 0$ in terms of
the known $\Phi_{00}$ and the unknown $\Phi_{22}$ we can regard $D_5=0$
as a (sub-linear) ordinary differential equation in $U$ for $\Phi_{22}$; hence, 
solving this equation delivers $\Phi_{22}$ and hence $\Phi_{11}$ and
$\Lambda$. 
Then, the condition $C_3=0$ produces $\Psi_2$.
Finally, after substituting for $\Psi_{2}$ and $\Phi_{11}$, the condition $D_{6}=0$
is a {\sl linear} equation for $\Psi_{4}$ with known data, and so 
we have determined a complete set of dependent variables on $\bNcal$.

When we consider the solution on the hypersurface $\Ncalb$ a
similar argument applies.
We see that $\Delta_{1}=\Delta_{2}=\Delta_{3}=\Delta_{4}=0$ is a coupled set of Riccati
ordinary differential equations for $\lambda, \mu$ 
with known sources and initial data on $\Pcal$. 
Using the data at $(U_0, V_0)$, we see that this differential system
has a unique solution  defined on a maximal interval 
$[V_0, \Vb_0)$. We then solve a linear system in $\rho, \sigma$. 
Using $C_{1}=0$ to express $\Phi_{11}\leqslant0$ in terms of the known
$\Phi_{22}$ and the unknown $\Phi_{00}$ we see that $\Delta_{5}=0$ is a
(sub-linear) equation for $\Phi_{00}$ with initial data on
$\Pcal$.  
This delivers $\Phi_{00}$ and hence $\Phi_{11}$ and $\Lambda$ on
$\Ncalb$ for a suitable finite $V$-interval.
Then $C_{3}=0$ produces $\Psi_{2}$.
Next $\Delta_{6}=0$ is a linear equation for $\Psi_{0}$ with known source
terms and given data at $\Pcal$, which delivers $\Psi_{0}$.
Finally $\Delta_{0}=0$ is a linear equation with known initial data for
$\epsilon$  and hence we have determined a complete set of dependent variables
on $\Ncalb$ for a suitable finite $V$-interval. 

The construction so far has produced solutions for $\epsilon$, $\rho$,
$\sigma$, $\lambda$, $\mu$, $\Phi_{00}$, $\Phi_{22}$, $\Psi_{0}$ and
$\Psi_{4}$ satisfying $D_{n}=0$ on $\bNcal$ and $\Delta_{n}=0$ on
$\Ncalb$. We have available 6 $D$-equations and 7 $\Delta$-equations to determine the
solution for these nine dependent variables in $\mathcal{M}$ the
future of $\bNcal\cup \Ncalb$.
We note that both $D_{1}=0$ and $\Delta_{1}=0$ could be used to evolve
$\rho$ and so we choose to discard one of them.  
As we shall see, it does not matter which.
We do the same for $\sigma$, $\lambda$ and $\mu$ dropping either
$D_{n}=0$ or $\Delta_{n}=0$ for $n=2, 3, 4$.
We now have {\sl $9$ equations for $9$ dependent variables.} 
We are not allowed to choose $a(u, v)$ freely in the interior,
although we know its value on $\Ncalb$.  
However one of the equations \eqref{C-metric} implies $D_{0}=0$ where
\be
  \label{eq:d0}
  D_{0}:= D a - \epsilon,
\ee
which we adjoin to our set, so that we have {\sl $10$ equations for $10$
independent variables} $a, \epsilon, \rho, \sigma, \lambda, \mu,
  \Phi_{00}, \Phi_{22}, \Psi_{0}, \Psi_{4}$, and there is precisely
  one $D$-equation or a $\Delta$-equation for each variable.
These form a first order quasilinear system, provided we use the
constraint equations $C_{k}=0$ to eliminate $\Phi_{11}$, $\Lambda$ and
$\Psi_{2}$. 
Further, writing the equations in the stated order it is obvious that
the system is diagonal and each entry has either a positive $l$-component or a positive $n$-component, 
and this guarantees that the solution exists and is unique.

The one remaining difficulty is that our system is {\sl not unique,} because
of the discarding process above---in fact there are $2^{4}$ such systems.
However, in each such system we set precisely four of the $\{D_{n}, \Delta_{n}\}$ ($n=1,2,3,4$) to zero 
(one for each choice of $n$), so there is no
\textit{a priori} guarantee that the others will also be zero.
We now examine the set \eqref{eq:allcom}, 
setting to zero all of the $C_{k}$ to zero, and those of the $D_{n}$ and $\Delta_{n}$ which we know to
be zero.
We also replace the coefficients in this set by the solution we
computed in the previous paragraph.
We are left with a linear symmetric hyperbolic system for the
remaining $D_{n}$ and $\Delta_{n}$ with zero source terms and trivial data
on $\bNcal$ or $\Ncalb$ as appropriate.  Clearly the
solution is the trivial one.
Thus the procedure described in this section generates 
a unique and complete set of dependent variables, at least for sufficiently regular solutions. 


\subsection{Weak regularity of the NP scalars}

The material in the present section provides a method to solve the characteristic initial
value problem and establish the existence of a solution within any characteristic rectangle 
$\Dcal(u_0, v_0; u,v)$ avoiding the singular line. 
Such a local existence result follows from standard theorems, in view of the 
symmetric hyperbolic form of the equations 
exhibited above. However, the above presentation does not provide a global existence result. 
In that sense, the analysis in Section~\ref{lowsec} 
was more precise and led us to an actual proof of existence, based on non-physical data though. 
Importantly, the present section has allowed us to identify the {\sl physically relevant data,}  
and, in the next section, we will put together our two approaches. 

It remains to discuss the regularity of the NP scalars when the spacetime is solely weakly regular.
By comparing the regularity in Definition~\ref{Hone} 
with the expressions of the Ricci and Weyl scalars derived in the present section, we arrive at the following regularity results.

\begin{proposition}
\label{theo-regul}
The Ricci and Weyl 
NP scalars associated with the weakly regular plane symmetric spacetimes 
defined in Theorem~\ref{F-Cauchy} within a characteristic rectangle $\Dcal$ have the following regularity
along the hypersurfaces $\bNcal$ and $\Ncalb$:  
\be
a \in W^{1,1}(\bNcal), \quad \Psi_0 \in W^{-1,2}(\bNcal), \quad \Phi_{00} \in L^1(\bNcal), 
  \label{C-dataN-regul}
\ee
and 
\be
a \in W^{1,1}(\Ncalb), \quad \Psi_4 \in W^{-1,2}(\Ncalb), \quad \Phi_{22} \in L^1(\Ncalb),
  \label{C-dataNdash-regul}
\ee 
as well as the following regularity in the spacetime: 
$$
\Phi_{11}, \Lambda, \Psi_2 \in L^1(\Dcal). 
$$
Moreover, these spacetimes satisfy $\Phi_{01} = \Phi_{02} = \Phi_{12} =0$ and $\Psi_1 = \Psi_3 =0$. 

\end{proposition}

This result provides us with a {\sl geometric} formulation of the regularity of $W^{1,2}$ weakly 
regular spacetimes. 

\begin{remark} {\sl In the coordinates chosen} in Theorem~\ref{F-Cauchy} the coefficient $b$ is
defined by \eqref {F-functionb2} and is thus smooth, but this regularity property is tight to our 
choice of characteristic coordinates. 
\end{remark}


\section{Global causal structure}
\label{glob} 

\subsection{Main result}   

This section is devoted to a proof of
a {\sl global and fully geometric} result which goes well 
beyond the local result given earlier in Theorem~\ref{F-Cauchy}.  
We still prescribe data on two null hypersurfaces intersecting along a two-plane, but
we are no longer working within a given characteristic rectangle and seek for the global structure of the 
future development of the given initial data set. 
Following the discussion in the 
previous section, we consider arbitrary null coordinates, denoted below by $(U,V)$, 
which in general differ from 
the coordinates $(u,v)$ constructed in Section~\ref{F-00}. 
Recall that, throughout, we restrict attention to {\sl plane symmetric} data and spacetimes. 

\begin{definition} {\sl 
An {\bf initial data with weak regularity}
consists of the following prescribed data.  
Let 
$$
\big(\bNcal, e^{\ba} dU dydz\big), \qquad \big(\Ncalb, e^{\ab} dV dydz\big)
$$ 
be two plane symmetric $3$-manifolds (endowed with volume forms) 
with boundaries identified along a two-plane $\Pcal$ and 
parametrized for some $(U_0, V_0)$ as  
$$
\bNcal := \big\{    U \geq U_0     \big\}, \qquad 
\Ncalb := \big\{    V \geq V_0     \big\}, 
\qquad   \Pcal := \left\{ U = U_0, \, V = V_0 \right\}.
$$ 
\begin{enumerate} 

\item Suppose that $\ba, \ab$ are absolutely continuous, 
i.e.~the integrals 
$$
\int_{\bNcal} \big( |\ba| + |\del_U \ba| \big) \,  e^{\ba} \, dU
\qquad 
\int_{\Ncalb} \big( |\ab| + |\del_V \ab| \big) \,  e^{\ab} \, dV
$$ 
are finite, and are normalized so that $\ba|_\Pcal = \ab|_\Pcal =0$. 
\item Let $\bPsi_0, \bPhi_{00}$ and $\Psib_4, \Phib_{22}$ be (plane-symmetric) functions 
defined on the hypersurfaces  $\bNcal$ and $\Ncalb$, respectively, with  
$0 \leq \bPhi_{00} \in L^1(\bNcal)$ and $0 \leq \Phib_{22} \in L^1(\Ncalb)$, i.e.~the integrals 
$$
\int_{\bNcal} \bPhi_{00} \,  e^{\ba} \, dU, 
\qquad 
\int_{\Ncalb} \Phib_{22} \,  e^{\ab} \, dV
$$ 
are finite, 
and that 
$\bPsi_0 \in W^{-1,2}(\bNcal)$ and $\Psib_4 \in W^{-1,2}(\Ncalb)$, i.e. $\bPsi_0 = \del_U \bPsi_0^{(1)}$
and $\Psib_4 = \del_V \Psib_4^{(1)}$
with 
$$
\int_{\bNcal} \big|\bPhi_{00}^{(1)}\big|^2 \,  e^{\ba} \, dU, 
\qquad 
\int_{\Ncalb} \Big| \Phib_{22}^{(1)}\big|^2 \,  e^{\ab} \, dV. 
$$ 

\item Finally, one also prescribes the connection NP scalars $\rho_0, \sigma_0, \lambda_0, \mu_0$ on $\Pcal$. 
\end{enumerate}  
}
\end{definition}

\ 

\begin{theorem}[Global causal structure of plane symmetric matter spacetimes]
\label{F-global} 
Consider an initial data set with weak regularity determined by
$\big(\bNcal, e^{\ba} \big)$ and $\big(\Ncalb, e^{\ab} \big)$, 
a plane $(\Pcal, \rho_0, \sigma_0, \lambda_0, \mu_0)$, 
and 
prescribed Ricci and Weyl NP scalars $\big(\bPsi_0, \bPhi_{00}\big)$ 
and $\big(\Psib_4, \Phib_{22}\big)$. 

(1) Then, there exists a unique $W^{1,2}$ regular spacetime $(\Mcal, g)$ determined by metric coefficients $a,b,c$
and matter potential $\psi$ which is a future development 
of the initial data set satisfying the Einstein equations \eqref{F-keyset} 
for self-gravitating, irrotational fluids with the initial conditions
\be
\label{onPcal}
(\rho, \sigma, \lambda, \mu) = (\rho_0, \sigma_0, \lambda_0, \mu_0) \quad \text{ on } \Pcal, 
\ee
\be
\big(a, \Psi_0, \Phi_{00}\big) =  \big(\ba, \bPsi_0, \bPhi_{00}\big) \quad     \text{ on the null 
hypersurface } \bNcal, 
  \label{C-dataN-bis}
\ee
and 
\be
\big( a, \Psi_4, \Phi_{22}\big) =  \big( \ab, \Psib_4, \Phib_{22}\big)  \quad \text{ on the null 
hypersurface } \Ncalb. 
  \label{C-dataNdash-bis}
\ee  

(2) The constructed development of the initial data
has past boundary 
$$
\big\{  \bU_0 >  U > U_0; V=V_0    \big\} \cup \big\{ U=U_0; \Vb_0 > V > V_0     \big\} \subset \bNcal \cup \Ncalb
$$
and, for {\bf generic initial data} (in the sense defined in (3), below) the curvature blows up to 
(and makes no sense even as a distribution) 
as one approaches  its future boundary
$$
\Bcal_0 := \big\{  F(U) + G(V) = 0     \big\}
$$ 
for some functions $F, G$ in $W^{1,2}$ (i.e.~having two derivatives in $L^1$), 
so that the spacetime is inextendible beyond $\Bcal_0$ within the class of $W^{1,2}$ regular spacetimes .

(3) This result holds for generic initial data, in the sense that arbitrary data can always be perturbed in the natural 
(energy-type) norm so that the perturbed initial data do generate a singular spacetime whose curvature blows-up on 
$\Bcal_0$. 

\end{theorem}

The coefficients $e^{\ba}$ and $e^{\ab}$, modulo a conformal transformation, could 
be chosen to be identically $1$ on the initial hypersurface, so that 
two main degrees of freedom remain on each of the two initial hypersurfaces. 
Theorem~\ref{F-global} can be seen as a statement of Penrose strong's censorship conjecture 
for plane symmetric spacetimes with low regularity.
The proof of this theorem requires a sufficient
 knowledge of the singularities of the Riemann function, which we discuss below.  
We emphasize that the theory developed in this paper applies to the matter model
\eqref{F-energy3} described at the end of Section~\ref{21}, which is equivalent to the
model \eqref{F-energy} as long as the energy density $w=\nabla^\alpha \psi \nabla_\alpha \psi$ remains positive. 


\subsection{Passage from metric data to NP scalar data} 

We first discuss the ``existence part'' in Theorem~\ref{F-global}. In principle, 
this result follows by applying   
Theorem~\ref{F-Cauchy} and patching together local solutions constructed in characteristic rectangles. 
The main difference between the two statements lies in the formulation of the initial data
and, thus, we need to check that the initial data posed on the NP scalars are sufficient to determine 
the initial data posed in terms of metric coefficients as was required earlier in Theorem~\ref{F-Cauchy}. 
 
We are given initial data in terms of conformal factors and curvature NP scalars prescribed 
on the two initial hypersurfaces, as stated in \eqref{C-dataN-bis}-\eqref{C-dataNdash-bis}. 
To recover the earlier description of the characteristic data in terms of the metric coefficients, we 
first observe that the two choices of null coordinates need not coincide.
So, we search for  
new characteristic variables
$$
u := F(U), \qquad v := G(V) 
$$
for some function $F,G,$ that remain to be identified and may {\sl depend on the prescribed data.} 
Roughly speaking, imposing $\Psi_0$ and $\Phi_{00}$ on $\bNcal$ is analogous to prescribing $c$ and $b$, 
respectively and similarly,  
imposing $\Psi_4$ and $\Phi_{22}$ on $\Ncalb$ is analogous to prescribing $c$ and $b$, respectively. 

Consider first the hypersurface $\bNcal$ on which we are given 
$$
\big(a, \Psi_0, \Phi_{00}\big) =  \big(\ba, \bPsi_0, \bPhi_{00}\big).  
$$
Our first task is, along $\bNcal$, to solve a Riccati matrix system in the variable $U$ for the unknown vector 
$(\brho, \bsigma):=(\rho, \sigma)|_{\bNcal}$. Using the quantities $D_1$ and $D_2$ introduced in \eqref{C-Dequations}
we find 
$$
\aligned
    & D \brho - \brho(\brho + 2\epsilon) - \bsigma^2 + \bPhi_{00} = 0, 
    \\
    & D \sigma - 2 \, (\rho + \epsilon) \bsigma - \bPsi_0 = 0, 
    \endaligned
$$
in which $\eps= a_U = \ba_U$ is a prescribed data along $\bNcal$, thus
$$
\aligned 
& D \begin{pmatrix} \brho \\ \bsigma \end{pmatrix} 
= \begin{pmatrix} \brho + 2 \ba_U & \bsigma 
\\ 2 \bsigma & 2 \ba_U \end{pmatrix} 
\begin{pmatrix} \brho \\ 
\bsigma \end{pmatrix} + \begin{pmatrix} \bPhi_{00} \\ \bPsi_0 \end{pmatrix} 
\quad 
& \text{ on } \bNcal. 
\endaligned
$$
The solution, in general, blows-up at some finite value denoted by $\bU_0$ which 
could be estimated by writing a matrix Riccati equation for  $P:=\begin{pmatrix} \brho & \bsigma 
\\ \bsigma &\brho  \end{pmatrix}$.  
Importantly, given the regularity of the Ricci scalar $\bPhi_{00} \in L^1$
and $\bPsi_0 \in W^{-1,2}$ we see that $\brho \in W^{1,1}(\bNcal)$ and $\bsigma \in L^2(\bNcal)$. 

Having identified the scalars $\brho, \bsigma$ on $\bNcal$ we can recover $\barb$ and $\bc$
by integration of \eqref{C-metric}: 
\be
\label{6666}
\barb(U) = - \int_{U_0}^U \brho(U') \, dU', 
\qquad 
\bc(U) = - \int_{U_0}^U \sigma(U') \, dU', 
\ee
where we have chosen the normalization $\barb(U_0) = \bc(U_0)=0$. Note that $\barb \in W^{2,1}(\bNcal)$
and $\bc \in W^{1,2}(\bNcal)$. We can then pursue the construction of the data on $\bNcal$ as was explained in Section~\ref{C-0}  
and, in particular, we recover the fluid potential as well. 

Similarly, along $\Ncalb$ the same argument produces the value $\Vb_0$ and the initial data $\bbar, \cb, \psib$. 

The analysis made earlier to show the existence of a weak solution within a characteristic rectangle applies
to show the existence of a solution $a,b,c, \psi$, where now 
{\sl the function $b$ is not normalized a~priori,}
 but at this stage $e^{2b}$ may be a general solution to the wave equation. 

The largest possible domain of interest is $[U_0, \Ub_0] \times [V_0, \Vb_0]$. However, not all of it is relevant since, in general, a blow-up in the function $b$ 
will take place before one can reach the boundary of this domain. 
The future boundary of the spacetime is determined by the function $b$, as follows. 
We determine the functions $F, G$ by considering $b$ along the initial hypersurfaces
$\bNcal$ and $\Ncalb$. Since $b$ is a geometric invariant (related to 
the area of the orbits of symmetry), it satisfies the wave equation derived earlier (in coordinates), 
and we can write 
$$
e^{2b(U,V)} = - \half F(U) - \half G(V) 
$$ 
for some functions $F,G$. These functions, when increasing, are used to define a change of (null) coordinates, 
defined by 
$$ 
u := F(U), \qquad v:= G(V). 
$$
In terms of the initial functions $\barb$ and $\bbar$ (already computed in \eqref{6666}),
 we find 
$$
F(U) = 2 \, e^{2\barb(U)} + G(V_0), \qquad G(V) = 2 \, e^{2\bbar(V)} + F(U_0),
$$
in which $F(U_0), G(V_0)$ are arbitrarily fixed with, since $\barb(U_0)=\bbar(V_0)=0$, 
$$
F(U_0) + G(V_0) = -2. 
$$ 
Finally, we apply Theorem~\ref{F-Cauchy} with suitable family of characteristic rectangles
covering the whole domain
$$
\Mcal:= \big\{  \bU_0 >  U > U_0; \Vb_0 > V > V_0; F(U) + G(V) < 0 \big\}.  
$$ 
To complete the proof of Theorem~\ref{F-global}, it remains to investigate the nature of the future boundary 
$$
\Bcal_0 := \big\{  F(U) + G(V) = 0     \big\} = \big\{ u+v = 0 \big\}. 
$$ 
 

\subsection{Blow-up behavior of the Riemann function} 
\label{BupEPD}
  
We now return to the notation introduced in Section~\ref{secEPD} within a characteristic rectangle, 
and reconsider the solutions $c, \psi$ constructed in Theorem~\ref{F-Cauchy}.
We need to investigate their behavior as the ``vertex'' $P=(u,v)$ of the characteristic rectangle approaches 
the singular line, i.e., $u+v\to 0-$. Recall that $z=0$ on $PQ$ and $PR$. Next, consider the line $QS$ and observe that $F(\half, \half; 1; z)=1$ at $Q$. In view of \eqref{eq:z}, 
when $u+v\to 0-$ one has also $z(S) \to 1-$.  

The Riemann function involves the function $F(\half, \half; 1; \cdot)$, which can be computed
thanks to the following \emph{Euler's formula} (valid for $|z| < 1$) 
\be
  \label{eq:Eulerform}
  \pi F(\half, \half; 1; z) = \int_{0}^{1} 
  \frac{dt}{t^{1/2}(1-t)^{1/2}(1-zt)^{1/2}},
\ee 
for which we refer to \cite{Olver} (Chapter~5, equation (9.01)) or \cite{AS65} (equation (15.3.1)). 

Clearly, as $z\to 1-$ the denominator in \eqref{eq:Eulerform} approaches
$t^{1/2}(1-t)$, leading to an integral which {\sl diverges logarithmically} 
at the upper end.
Thus, for $z$ close to $1$, the dominant contribution to the integral comes from the upper end, 
and we can quantify this property as follows. By defining the function 
$$
  I(t,z) = \int^t_0 \frac{d\tau}{(1-\tau)^{1/2}(1-z\tau)^{1/2}}, 
$$
we can check that, at the leading order, 
$$
  \pi F(\half, \half; 1; z) \sim I(1,z)  \quad \text{ as } \, z\to 1-. 
$$ 
On the other hand, the integral $I(1,z)$ can be computed explicitly and we conclude 
\be
  \label{eq:Fasym}
   F(\half, \half; 1; z)\sim
   -\frac{1}{\pi}\log(1-z)\quad
   \text{ as }z\to 1-.
\ee

We return to the representation \eqref{eq:crg10}, and note first that
in each of the integrands, the Riemann function $\varphi$ has constant sign so that no cancellation can
take place within each term.   Consider, for instance, 
the first integral (over the segment $SQ$) and assume that $\barB[\psi](\cdot, v_0)$ is a ``generic'' function
so that the behavior of the first integral term in \eqref{eq:crg10} can be determined by studying 
\be
  \label{eq:phi1}
 f(u,v) : = \int_{u_0}^{u} 
  \left(\frac{u'+v_0}{u'+v}\right)^{1/2}
  \left(\frac{u'+v_0}{u+v_0}\right)^{1/2}
  F\left(\half, \half; 1; \frac{(v'-v)(u'-u)}{(v'+u)(u'+v)}\right) \, du',
\ee
in which we have specified the relevant value of the argument $z$. 

In the region $u+v<0$ the function $f$ is regular, and we are interested here in the limit 
$$
  u+v := -\epsilon \to 0-,
$$
in which the coefficients of the Euler-Poisson-Darboux equation \eqref{eq:crg1} becomes singular.
Clearly, a singularity in $f(u,v)$ must arise at one (or both) endpoint(s) of the integral.

Near $Q$, $u'\approx u$, $z\approx 0$, and both the second and third
factors in the integrand \eqref{eq:phi1} are approximately unity.
Then $f(u,v)$ picks up a term $|u+v_0|^{1/2}|u+v|^{1/2}$ which is
itself finite in the limit $- \eps=u+v \to 0$. 
However its first derivatives are $O(\epsilon^{-1/2})$,  
singular as $\epsilon\to 0$. 

Near $S$, $u'\approx u_0$ and we see from \eqref{eq:z} that
$1-z=O(\epsilon)$ and so 
$F(\half, \half; 1; z)\sim -\pi^{-1}\log \epsilon$ and
$f(u,v)\sim -K\log \eps$ where $K$ is a strictly positive constant.
Not only does $f(u,v)$ become singular when $- \eps=u+v \to 0$, 
but its first derivatives are $O(\epsilon^{-1})$.  
This behavior dominates the milder singularity which originates from
a neighborhood of $Q$.

Exactly the same analysis, with exactly the same result, can be
applied to the second integral (over $SR$) in \eqref{eq:crg10}.
We have 
$$
	\varphi(u_0, v_0; u,v) \sim -K \, \log|u + v| \qquad \text{ as } - \eps= u+v \to 0- 
$$
for some $K$. 
    Thus in this limit there are three terms in \eqref{eq:crg10} that will pick
    up a $\log(u+v)$ factor, the first term, and a contribution from
    the lower end of both integrals.  The magnitude of these terms is
    determined by the behaviour of the characteristic initial data 
and, for generic data, 
    cancellation will not occur.

Based on the above observations we can write the principal part of a general solution
\be
  \label{ak9}
  \aligned
   \psi(u,v) = & T(u,v) + \bT(u,v) + \Tb(u,v)
    \\
    T(u,v) = & \varphi(u_0,v_0;u, v) \, \psi(u_0, v_0),   
   \\
   \bT(u,v) = & \int_{u_0}^{u}\varphi(u',v_0;u,v) \, \barB[\psi](u', v_0) \, du', 
     \\
   \Tb(u,v) = & \int_{v_0}^{v}\varphi(u_0,v';u,v) \, \Bbar[\psi](u_0, v'] \, dv',
  \endaligned
\ee
with 
\be
\label{ak10}
\aligned
\barB[\psi](u', v_0)  :=& \psi_u(u',v_0) + \half (u'+v_0)^{-1}\psi(u',v_0), 
\\
\Bbar[\psi](u_0, v'] :=& \psi_v(u_0,v') + \half (u_0+v')^{-1}\psi(u_0,v'). 
\endaligned
\ee 
and 
$$
\varphi(u',v';u,v) 
=
  \left(\frac{u'+v'}{u'+v}\right)^{1/2}
  \left(\frac{u'+v'}{u+v'}\right)^{1/2}
  \Ft\left( \frac{(u'+v')(u+v)}{(u'+v)(v'+u)}\right), 
$$ 
where for simplicity we have set $\Ft(y) := F(\half, \half; 1, 1-y)$. 

We find
$$
\aligned
T(u,v) = & \psi(u_0, v_0) \,   \left(\frac{u_0+v_0}{u_0+v}\right)^{1/2}
  \left(\frac{u_0+v_0}{u+v_0}\right)^{1/2}
  \Ft\left(  \frac{(u_0+v_0)(u+v)}{(u_0+v)(v_0+u)}\right)
\\
\sim & -{1 \over \pi} \, {(u_0+v_0) \, \psi(u_0, v_0) \over (u_0-u)^{1/2} \, (u+v_0)^{1/2}} \, \log |u+v|
\\
=: & T(u) \, \log |u+v|
\endaligned
$$
and, for the first integral term,  
$$
\aligned
& \bT(u,v) 
\\
& = \int_{u_0}^{u} \barB[\psi](u', v_0) \, 
  \left(\frac{u' + v_0}{u' -u + \eps}\right)^{1/2}
  \left(\frac{u' + v_0}{u + v_0}\right)^{1/2}
  \Ft\left( \frac{(u' + v_0) \eps}{(u' - u + \eps) (v_0 + u)}\right) \, 
  du'
  \\ 
& \sim -{1 \over \pi} \, \int_{u_0}^{u} \barB[\psi](u', v_0) \, 
  \left(\frac{u' + v_0}{u' - u}\right)^{1/2}
  \left(\frac{u' + v_0}{u + v_0}\right)^{1/2} \, du' \, \log |u+v|
  \\
& =: \bT(u) \, \log |u+v|. 
\endaligned
$$
The calculation for $\Tb(u,v)$ is similar.

This establishes that the leading term in the asymptotic expansion of the function $\psi$ is of the form
\be
  \label{eq:asy1} 
  \psi(u,v) \sim \Psi(u) \, \log |u+v| \qquad  \text{ as } -\eps = u+v \to 0, 
\ee
and similarly for the solution $c$ 
\be
  \label{eq:asy12}
  c(u,v) \sim C(u) \, \log |u+v| \qquad  \text{ as } -\eps = u+v \to 0, 
\ee  
where $C$ and $\Psi$ are certain functions that are given explicitly by the above formulas and 
are generically non-vanishing. These coefficients make sense as distributions: for instance $\bT(u)$ 
is defined the limit of the sequence
\be
\label{xyz}
\bT^\eps(u) := \int_{u_0}^{u} \barB[\psi](u', v_0) \, 
  \left(\frac{u' + v_0}{u' -u + \eps}\right)^{1/2}
  \left(\frac{u' + v_0}{u + v_0}\right)^{1/2} \, du'. 
\ee
Each term $\bT^\eps(\cdot)$ belong to $L^2$ (as follows from the regularity assumed on the data)
and, as a sequence, converges to the distribution $\bT(\cdot)$.  Hence, for both $C$ and $\Psi$ we 
have the existence of $C^\eps, \Psi^\eps$ such that 
\be
\label{reg8}
(C, \Psi) = \lim_{\eps \to 0} (C^\eps,\Psi^\eps)(\cdot), \qquad (C^\eps,\Psi^\eps)(\cdot) \in L^2, 
\ee 
where the convergence holds in the distributional sense. 


\subsection{Proof of the main result}
\label{54section}

\subsubsection*{Blow-up analysis}

We may infer the leading term in the asymptotic expansion of the
metric function $a$ from \eqref{F-keyset} as
\be
  \label{eq:asy2}
  a(u,v) \sim A(u) \, \log |u+v|, \qquad A = C^{2} + \half \Psi^{2}- \quarter 
  \geq -\quarter. 
\ee
More precisely, one has 
\be
\label{reg8-5}
A=\lim_{\eps \to 0} A^\eps, \qquad A^\eps \in L^1,
\ee 
where the convergence holds in the distributional sense. 
The behavior of the metric function $b$ is given by \eqref{F-functionb2} as
\be
  \label{eq:asy3}
  b \sim \half\log |u+v|,
\ee
and so we can construct the leading terms in the asymptotic
expansion of the metric
\be 
  \begin{aligned} 
    g \sim |u + v|^{2A} \, du dv - |u + v|^{1+2C} \, dy^2 + |u + v|^{1-2C} \, dz^2 \big).  
  \end{aligned}
  \label{F-metric55}
\ee

The Ricci and Weyl curvature tensors are best described using the
Newman-Penrose formalism and explicit formulae for the NP scalars
$\Phi_{mn}$, $\Lambda$ and $\Psi_{n}$ were given in Section~\ref{C-0}. 
After some calculations, the Ricci and Weyl invariants as $|u+v| \to 0$ are given by
$$
    R = 24\Lambda\sim \quarter(4C^{2}-4A-1) \, |u + v|^{-2-2A}, 
$$
$$
  \begin{aligned}
\big|  R_{\alpha\beta}R^{\alpha\beta} \big|^{1/2}
&= 
    2 \sqrt{2} \, \big| \Phi_{00}\Phi_{22} + 18\Lambda^{2}+2\Phi_{11}{}^{2} -
    4|\Phi_{01}|^{2} + |\Phi_{02}|^{2} \big|^{1/2}
    \\
    & \sim \tfrac{1}{4 \sqrt{2}} \, \big| 4A-4C^{2}+1\big| \, (1+ |u + v|^{-4A}\big)^{1/2} \, |u + v|^{-2}, 
  \end{aligned}
$$
and
$$
  \begin{aligned}
   \big| C_{\alpha\beta\gamma\delta}C^{\alpha\beta\gamma\delta} \big|^{1/2}
   &=
    4 \, \big| 3\Psi_{2}{}^{2}+\Psi_{0}\Psi_{4} - 4\Psi_{1}\Psi_{3} \big|^{1/2} 
    \\
    & \sim \Big| \tfrac{1}{12}(2A+4C^{2}-1)^2 \, |u + v|^{-4A} 
    + 4 \, A^{2}C^{2} \, |u + v|^{4A} \Big|^{1/2} \, |u + v|^{-2}.
  \end{aligned}
$$
We emphasize that $\Phi_{00}, \Phi_{01}, \Phi_{11}, \Phi_{22}$ belong to $L^2$ so that the scalar 
$\big(    R_{\alpha\beta}R^{\alpha\beta} \big)^{1/2}$ belong to $L^1$. On the other hand, the Ricci scalar 
has an expansion whose the principal terms's coefficient is the limit of functions in $L^1$. 

A necessary condition to avoid the  
blowing up of the curvature of the spacetime at $u+v=0$ 
is that these
leading terms vanish, which occurs if and only if 
\be
\label{BUP}
C^{2}- \quarter =  A = 0. 
\ee
This condition implies $\Psi =0$. 
Even if these very special conditions are fulfilled there is no
guarantee that spacetime would not blow-up---one would have to investigate the next
order terms. We conclude  
that in the case when the above coefficients do not vanish, $u+v=0$ is a strong
curvature singularity. 

Finally, we emphasize that by perturbation of the data in the natural energy norm (corresponding to their assumed regularity), we can always ensure that the above coefficient do not vanish. Indeed, if \eqref{BUP} holds, 
then by adding a constant $\alpha$ to the characteristic initial data for $\psi$, we see that 
$$
\aligned
\barB[\psi + \alpha](u', v_0)  :=& \barB[\psi](u', v_0) + \half (u'+v_0)^{-1}\alpha, 
\endaligned
$$
so that the coefficient on the singularity (given by the limit of \eqref{xyz}) 
is changing accordingly by a (non-vanishing) term proportional 
to $\alpha$. In turn, $C^{2}- \quarter$ and $A$ must be non-vanishing for all sufficiently small $\alpha$, at least. 
This completes the proof of Theorem~\ref{F-global}. 
 

\subsection{Special solutions}

It is of interest to consider the special case of trivial initial data, that is, when there are no gravitational waves 
and no matter content. For simplicity let us set $u_0=v_0=0$. As noted earlier, $r= e^{b}$ satisfies 
$\del_u \del_v (r^2) = 0$ nd, after normalization, $r^2 = f(u)^2 + g(v)^2 -1$ with 
$f(u) = r(u,0)$, $g(v)=r(0,v)$, and $(f(0) = g(0) = r(0,0)=1$. 

When the free data vanish, we can solve the relevant Einstein equations and obtain that $f$ and  $g$ vary linearly, that is, $f(u)=1+ \alpha u$ and $g(v)=1+ \beta v$ for some constants $\alpha, \beta$. 
We distinguish between two cases whether these constants have the same sign or opposite signs and
we make here specific sign choices since the other cases can be deduced by time or space reversals. In addition, 
by taking advantage of the transformation $(u,v) \mapsto (pu, v/p)$ we can finally consider 
Case 1: $\alpha=\beta >0$ and 
Case 2: $-\alpha =\beta >0$. 

With vanishing free data, the metric reads $g= e^{2a} (du^2 + dv^2) - r^2 (dy^2 + dz^2)$, 
while the contraint \eqref{eq:C-algebraic1} takes the form $\Psi_2 = e^{-2a} r^{-2} \del_u r \del_v r$
and the Bianchi identities become $\del_u (r^3 \Psi_2)= \del_v (r^3 \Psi_2) = 0$. Hence, for some constant 
$m$ we have $r^3 \Psi_2 = -2m$. 

Consider first Case 1 above. Along $\bNcal$ one has $\Psi_2 = r^{-2} \del_u r \del_v r$ and 
$$
r= 1 + \alpha u, \qquad r \del_u r = \alpha (1 + \alpha u), \qquad 
r \del_v r = \alpha. 
$$
Therefore, $r^3 \rho = -\alpha^2$ and $2m = \alpha^2>0$ and we obtain $e^{2a} = (1+ \alpha u) (1 + \alpha v)/r$. 
In this case, we have $u > -1/\alpha$ and $v > -1/\alpha$. 
Coordinate singularities arise at $u= -1/\alpha$ where 
$\del_u r$ vanishes, as well as at $v= -1/\alpha$ where 
$\del_v r$ vanishes. More importantly, a genuine (geometric) singularity arises at $r=0$, that is, on the ``quarter circle'' 
determined by the conditions 
$$
(u+ 1/\alpha)^2 + (v+ 1/\alpha)^2 = 1/\alpha^2, 
\qquad 
u > -1/\alpha, 
\quad v > -1/\alpha.
$$

If now we make a conformal transformation of the metric we can set $\alpha =1$ and apply the transformation 
$u\mapsto u-1$ and $v \mapsto v-1$, so that the two metric components are given simply by 
$$
e^{2a} = {uv \over r}, \qquad  r^2 = u^2 + v^2 -1. 
$$
The quotient manifold $Q$ then corresponds to the part of the positive quadrant which lies outside the unit circle. 
The maximal extension is obtained by introducing $U= u^2/2$ and $V = v^2/2$ which removes the coordinate singularity. By setting $U= \tau - x$ and $V=\tau +x$ the metric becomes 
$$
g = (4\tau-1)^{-1/2} (d\tau^2-dx^2) - (4\tau -1) (dy^2 + dz^2), 
$$
and the extension covers the domain $\tau >1/4$. Finally, setting $3t = (4\tau - 1)^{3/4}$ and making the transformation $(x,y,z) \mapsto (3^{1/3} x, 3^{-2/3} y, 3^{-2/3} z)$, we see tha the metric takes the well-known 
Kasner form
$$
g= dt^2 - t^{-2/3} dx^2 - t^{4/3}  (dy^2 + dz^2). 
$$
In addition to the assumed symmetry made for general solutions, this special case enjoys a rotational symmetry in the $(y,z)$-plane and, furthermore, the Kasner metric is easily checked to be conformally invariant so that the constant 
$\alpha$ has been absorbed and we arrive to only one geometrically distinct solution rather than a one-parameter family of solution as one could have expected.  

Concerning Case 2 mentioned earlier, a very similar analysis can be performed which leads us to 
the metric expression 
$$
g= x^{-2/3} dt^2 - t^{-2/3} dx^2 - x^{4/3} (dy^2 + dz^2). 
$$
This is now a static solution with a timelike singular boundary. Again, this metric is invariant under homotheties, so that 
the constant $\beta$ is absorbed and there is again only one geometrically distinct solution.


\subsection{An alternative approach} 

Another approach to studying the behavior of solutions on the coordinate singularity 
is provided by imposing data directly on the singularity hypersurface $u+v=0$, 
as we now discuss.  
Following Hauser and Ernst~\cite{HE}, 
we start from the general form of an EPD equation 
\begin{equation}
  \label{eq:1}
  (u+v)\psi_{uv} + \alpha\psi_{u}+\beta\psi_{v}=0,
\end{equation}
where $\alpha,\beta \in (0,1)$ are constants.
It is easy to check that $(u-\sigma)^{-\beta}(v+\sigma)^{-\alpha}$ is a solution
for any constant $\sigma$. Then, superposing such solutions we see that
\begin{equation}
\label{eq:20}
  \psi(u, v) = \int_{X}^{Y}\frac{A(\sigma)\, d\sigma}{(u-\sigma)^{\beta}(v+\sigma)^{\alpha}}
\end{equation}
is also a solution, where $A=A(\sigma)$ is an arbitrary function and $X$
and $Y$  are constants. A formula due to Poisson allows also to take $X=u$ and $Y=-v$. 
Hauser and Ernst looked at the case $X=u_{0}, Y=u$ with
$\alpha=\beta=\half$, and observed that
\begin{equation}
  \label{eq:2}
  \psi(u, v) =
  \int_{u_{0}}^{u}\frac{A(\sigma)\,d\sigma}
  {\sqrt{u-\sigma}\sqrt{v+\sigma}} = 
  \int_{u_{0}}^{u}\frac{\widehat{A}(\sigma)\sqrt{v_{0}+\sigma}\,d\sigma}
  {\sqrt{u-\sigma}\sqrt{v+\sigma}}
\end{equation}
is a solution of \eqref{eq:1} with $\alpha=\beta=\half$.

For the moment we accept this assertion, and evaluate the solution
when $v=v_{0}$,
\begin{equation}
  \label{eq:5}
  \psi(u,v_{0}) = \int_{u_{0}}^{u}\frac{\widehat{A}(\sigma)\,d\sigma}{\sqrt{u-\sigma}}.
\end{equation}
Now the range of this integral is the line $SQ$, 
and the
left hand side is (that half of) the characteristic initial data which
we wish to impose on $SQ$.
Note that we are assuming $\psi(u_{0},v_{0}) = 0$.
If $\psi$ represents the velocity potential this is not a problem.
If $\psi$ represents the metric coefficient $c$ we may need to scale
the ignorable coordinates $y$ and $z$ to achieve this.

As Hauser and Ernst pointed out, \eqref{eq:5} is an Abel integral
equation which can be solved explicitly for $\widehat{A}(\sigma)$: 
\begin{equation}
  \label{eq:6}
  \widehat{A}(\sigma) = \frac{1}{\pi}\frac{\partial}{\partial \sigma}
    \int_{u_{0}}^{\sigma}\frac{\psi(u,v_{0})\,du}{\sqrt{\sigma-u}}.
\ee 
Since $\alpha=\beta$, we can repeat the argument by interchanging $u$ and $v$ and this leads us to the formula 
\begin{equation}
  \label{eq:7}
  \psi(u, v) =
  \int_{v_{0}}^{v}\frac{\widehat{B}(\sigma)\sqrt{u_{0}+\sigma}\,d\sigma}
  {\sqrt{v-\sigma}\sqrt{u+\sigma}},\qquad
  \widehat{B}(\sigma) = \frac{1}{\pi}\frac{\partial}{\partial \sigma}
    \int_{v_{0}}^{\sigma}\frac{\psi(u_{0},v)\,dv}{\sqrt{\sigma-v}},
\end{equation}
produces a solution of the EPD equation which fits the characteristic
initial data on $SR$. 
By linearity, the general solution to the 
characteristic initial value problem
is then given by summing \eqref{eq:2} and \eqref{eq:7}. We thus have obtained 
explicitly the dependence of the solution on the initial data. 

We do need to check that \eqref{eq:2} and \eqref{eq:7} are in
fact solutions which can be continued up to the singular line $u+v=0$.
(This is not a problem for the Poisson-Appell version \eqref{eq:20}.)
Concentrating on \eqref{eq:2} we change the variable of integration so
that the integral is over a fixed range by setting
$$
\aligned
&  \sigma=(1-\tau)u_{0}+\tau u,
\\
& f_{0}= u-u_{0},\qquad
  f_{1}= 1-\tau,\qquad f_{2}= v+u_{0}+ \tau(u-u_{0}).
\endaligned
$$
Then Hauser and Ernst assert that
\begin{equation}
  \label{eq:4}
  \psi(u,v) = \int_{0}^{1}\frac{A(\sigma)\sqrt{f_{0}}\,d\tau}
  {\sqrt{f_{1}}\sqrt{f_{2}}}
\end{equation}
is a solution for arbitrary $A(\sigma)$.
Suppose we define
\begin{equation}
  \label{eq:9}
  G(\tau,u,v) = \frac{A(\sigma)(f_{2}-u-v)\tau}
  {\sqrt{f_{0}}f_{1}\sqrt{f_{1}}f_{2}\sqrt{f_{2}}}.
\end{equation}
Then it is straightforward to verify that \eqref{eq:4} is indeed a
solution for arbitrary $A(\sigma)$ provided the ``surface term''
\begin{equation}
  \label{eq:10}
  \bigg[G(\tau,u,v)\bigg]_{\tau=0}^{\tau=1}
\end{equation}
vanishes.
This is a delicate matter.
Near $\tau=0$
\begin{equation*}
  G(\tau,u,v)\sim -\,\frac{A(u_{0})\sqrt{u-u_{0}}\,\tau}{(v+u_{0})\sqrt{v+u_{0}}}\,,
\end{equation*}
and this vanishes as $\tau\downarrow 0+$ provided $v+u_{0}\ne 0$.
Near $\tau=1$
\begin{equation*}
  G(\tau,u,v) \sim -\,\frac{A(u)\sqrt{u-u_{0}}\sqrt{1-\tau}}
  {(u+v)\sqrt{u+v}}\,.
\end{equation*}
Away from the singular line $u+v<\epsilon<0$, the limit as
$\tau\uparrow 1-$ is straightforward, and \eqref{eq:2} does indeed
furnish a solution of the EPD equation.
However if we want to evaluate $\psi(u,v)$ near the singular line,
$(u+v)\uparrow 0-$, the (somewhat ill-defined) order in which we take
the limits is crucial.

Suppose we accept that \eqref{eq:2} is a solution.  In general we
cannot calculate its value. 
As a special case suppose that $A(\sigma)$ is a constant, say $ia$.
(The $i$ factor  allows for the fact that $v+\sigma < 0$ in our
scenario.)
The indefinite integral is easily seen to be
\begin{equation*}
  a\log(\sqrt{u-\sigma}\sqrt{-v-\sigma} + 2\sigma - u + v),
\end{equation*}
and 
\begin{equation}
  \label{eq:11}
  \psi(u, v) = a\left(\log(u+v) - \log(2\sqrt{u-u_{0}}\sqrt{-v-u_{0}}
    + 2u_{0}- u + v)\right),
\end{equation}
which exhibits the $\log(u+v)$ singularity that we expect. 
We conclude with the following question: The condition for avoiding the log singularity is that
$A(\sigma)\to 0$ sufficiently fast as $\sigma\uparrow u-$.
Clearly $\widehat{A}(\sigma)$ must behave in the same way. It would be interesting to 
use \eqref{eq:6} and relate this to the behavior of
$\psi(u, v_{0})$. 

 
Let us make a final remark when analyticity of the data on the singularity is assumed.  
We recast the EPD equation
\eqref{eq:crg1} in the form  
\be
  \label{eq:sing1}
  L[\psi] = M[\psi],
\ee
where $L[\psi] = \psi_{tt} + \psi_{t}/t$ and $M[\psi] = \psi_{xx}$, 
e.g., $L[t^n] = n^{2}t^{n-2}$, $M[x^{n}] = n(n-1)x^{n-2}$.
Suppose we have a sequence of functions $\{\ldots, \omega_{n}(t),
\ldots\}$ with the property $L[\omega_{n}] = \omega_{n-1}$,
and a sequence of functions $\{\ldots, \phi_{n}(x),\ldots\}$ with the
property $M[\phi_{n}] = \phi_{n+1}$. 
Now, ignoring all questions of convergence, it is easy to see that
$\psi(t,x) = \sum_{m=-\infty}^{\infty}\omega_{m}(t)\phi_{m}(x)$
satisfies the EPD equation.

On physical grounds we do not expect $\psi$ to contain arbitrarily
large negative powers of $t$ as $t\to 0$ and so we require
$\omega_{n}(t)=0$ for all $n < n_{0}$.
Since $m$ in the sum above is only defined up to an arbitrary additive
constant we may set $n_{0}=0$, i.e.,
$$
  \psi(t, x) = \sum_{n=0}^{\infty}\omega_{n}(t)\phi_{n}(x).
$$
Clearly $L[\omega_{0}]=0$ which implies
$\omega_{0}(t) = A + B\log t$, 
where $A$ and $B$ are constants.
This in turn demands
$$
  \omega_{n}(t) = A(a_{n}t^{2n}) + B(a_{n}t^{2n}\log t + b_{n}t^{2n}),
$$
where $a_{n}= 2^{-2n}(n!)^{-2}$ and $b_{n}$ is defined by a recurrence
relation
$$
  b_{n}= (2n)^{-2}(b_{n-1}-4na_{n})
$$
with $b_{0}=0$. 
Suppose that at $t=0$ the solution of the EPD equation is
\be
  \label{eq:singdat}
  \psi(t,x) = A\alpha(x) + B \beta(x)\log t.
\ee
Then, within the class of real analytic solutions one finds 
\be
  \label{eq:singsol}
  \psi(t,x) = A\sum_{0}^{\infty}a_{n}t^{2n}\alpha_{n}(x) + 
  B\sum_{0}^{\infty}(a_{n}\log t + b_{n})t^{2n}\beta_{n}(x),
\ee
where
$$
  \alpha_{n}(x) = \left(\frac{d\;\,}{dx}\right)^{2n}\alpha(x), 
  \qquad
  \beta_{n}(x) = \left(\frac{d\;\,}{dx}\right)^{2n}\beta(x).
$$
For instance, the exact solution cited in Remark~\ref{rem31} is generated by $A=8$, $B=0$, and $\alpha(x)=x^2$. 

Clearly $\psi$ is regular at $t=0$ if and only if $B=0$, and such a
solution will give a regular $\psi$ and $\psi_{t}$ on another Cauchy
surface, say $t=t_0\ne0$. 
However any slight change to the data at $t=t_{0}$ means that $B$
changes away from zero, and so the resulting solution will become
singular as $t\to 0$.
We deduce that generic solutions of a Cauchy problem blow up as
$t\to 0$. 


\section{Propagation of curvature singularities}
\label{S4}

\subsection{Choice of tetrad and general expression of the metric}

In this section we determine which kind of curvature singularity is allowed on
a null hypersurface, denoted below by $\Ncal_0$, for solutions of the Einstein equations. 
No symmetry assumption is made in our general discussion. However, in each statement below, we 
specify what simplification is achieved for plane symmetric spacetimes. Hence, this section provides 
us a geometric derivation of jump relations satisfied by the spacetimes constructed in the rest of this paper, 
which could also be established from the distributional formulation of the Einstein equations that we 
have discussed. 

A main source of difficulty comes from the fact  that the metric
induced on a null hypersurface (from the spacetime metric) 
is {\sl degenerate,} with signature $(0, -1, -1)$.  
In consequence, several fundamental concepts (linear connection, etc)
must be re-visited. 
This was done first by Stellmacher \cite{Stellmacher}, but we 
adopt here the more geometrical approach due to  Penrose \cite{Penrose}.
He pointed out that it is natural to distinguish between three
distinct types of geometries on a null hypersurface and, in consequence,
to impose the condition that these geometries coincide when taking
the limit of the spacetime metric from either side of the
hypersurface.  

We may distinguish between the following three classes of spacetimes $\Mcal= \Mcal^- \cup \Mcal^+$ 
singular across a null hypersurface $\Ncal_0 := \Mcal^- \cap \Mcal^+$: 
\begin{enumerate}

\item[$\bullet$] I-Geometry : the induced, degenerate metric only
  is continuous across $\Ncal_0$.  
  This is the most general situation where most curvature scalars contain Dirac measure terms.  

\item[$\bullet$] II-Geometry : a concept of parallel transport, along each 
 integral curve $\gamma$, of tangent vectors to $\gamma$ is assumed to coincide on both sides of the hypersurface. 

\item[$\bullet$] III-Geometry : a concept of parallel transport, along
  each integral curve $\gamma$, of tangent vectors to the hypersurface $\Ncal$ is now assumed.  
\end{enumerate}

The original presentation by Penrose was based on the use of $2$-component spinors.
Our presentation will use the more familiar Newman-Penrose formalism
based on a null tetrad, which avoids the explicit use of spinors, but restricts our
consideration to II- and III-geometries. 
We will construct the spacetime by starting from an given arbitrary spacelike two-surface 
and considering hypersurfaces generated by the two  families of null geodesics
orthogonal to that two-surface. For each singular geometry, 
we will then investigate which singularities are admissible along these hypersurfaces. 

For the moment we assume that all of the quantities under consideration
are sufficiently regular. 
Let $\Pcal_{0,0}$ be an arbitrary spacelike two-surface with intrinsic
coordinates $(x^A)$ with $A = 1,2$. 
Through each point $P_{0,0} \in \Pcal_{0,0}$ there pass two null geodesics
orthogonal to the tangent space $T \Pcal_{0,0}$.  
As $P_{0,0}$ varies (and provided they do not recross), these geodesics 
trace out two null hypersurfaces which we denote by $\Ncal_0$ and $\Ncal_0'$,
respectively. Let $u$ be a parameter along the generators of $\Ncal_0$, normalized so that 
$$
u = 0 \quad \text{ on } \Pcal_{0,0}.
$$ 
The parameter $v$ is similarly defined on $\Ncal_0'$, and neither
parameter need be affine.  

Let $\Pcal_{u,0}$ be the (necessarily spacelike) two-surface determined by
the condition $u=$constant in $\Ncal_0$. 
Through each point $P_{u,0} \in \Pcal_{u,0}$ there passes a second null geodesic
orthogonal to $T \Pcal_{u,0}$ and (assuming again that no conjugate point arises),
these geodesics trace out another null hypersurface $\Ncal_u'$. 
We extend the definition of the parameter $u$
by requiring that $u$ remains constant on $\Ncal_u'$.  
In a similar way we construct a family of null hypersurfaces $\Ncal_v$ and, 
now, $u$ and $v$ are two of the four coordinates on the spacetime $\Mcal$ which, without
loss of generality, we assume 
to be entirely covered by our construction. The
two-surfaces on which $u$ and $v$ remain constants is denoted by $\Pcal_{u,v}$. 

We extend the definition of $(x^A)$ from $\Pcal_{0,0}$ to $\Pcal_{u.0}$
by requiring that $x^A$ be constant along the generators of $\Ncal_0$. 
We then extend the definition of $(x^A)$ to the surface $\Pcal_{u,v}$ by
requiring that $x^A$ is constant along the generators of $\Ncal_u'$.  
Note that, in general, $x^A$ is {\sl not constant} along the generators
of $\Ncal_v$, the exception being the initial hypersurface $\Ncal_0$.   

Importantly, we can carry out these constructions both for $v>0$ and for 
$v<0$, and these are (locally) distinct and lead to two pieces of our spacetime, 
denoted by $\Mcal^+$ and $\Mcal^-$, respectively.
By construction, the coordinate charts are continuous at $\Ncal_0 = \Mcal^+ \cap \Mcal^-$, 
but no further regularity is available in general.

Consider, for instance, the manifold (with boundary) $\Mcal^-$. 
The surfaces $\Ncal_u' \cap \Mcal^+$ are null hypersurfaces in $\Mcal^+$ on which we have the null covector field  
$$ 
n_\alpha \, dx^\alpha = du, \quad \mbox{ or equivalently } \quad  n_\alpha = \nabla_\alpha u
\quad \mbox { in } \Mcal^+,
$$  
and the vector field $n^\alpha$ is parallel to the generators of $\Ncal_u'$, along which only 
$v$ varies. Thus, there exists a scalar function $Q >0$ such that 
\be 
\Delta := n^\alpha \, {\del \over \del x^\alpha} = Q \, {\del \over \del v} \quad \mbox{ in } \Mcal^+. 
\label{eq : 4.a2}
\ee 

Similarly, the surfaces $\Ncal_v \cap \Mcal^+$  are null hypersurfaces on which we can define
the null covector field  
$$ 
l_\alpha \, dx^\alpha = Q^{-1} \, dv \quad \mbox{ in } \Mcal^+, 
$$
which enforces the condition $l_{a}  n^\alpha = 1$ in $\Mcal^+$. The vector $l^\alpha$ 
is tangent to the generators of $\Ncal_v$ along which $u$ and $x^A$ change. 
Enforcing $n_\alpha \, l^\alpha = 1$ we deduce that 
\be 
D := l^\alpha \, {\del \over \del x^\alpha} = {\del \over \del u} + C^A \,
{\del \over \del x^A} \quad  \mbox{ in } \Mcal^+,
\label{eq : 4.a4}
\ee 
where $C^A$ are two real-valued, scalar functions. 

Finally, consider the tetrad vectors $m^\alpha$ and $\om^\alpha$. 
Since $n_{a}m^\alpha = l_{a}m^{a} = 0$ by definition, they cannot have
$u$ and $v$-components, and we thus set 
\be 
\delta := m^\alpha \, {\del \over \del x^\alpha} = P^A \, {\del \over \del x^A} 
\quad \mbox{ in } \Mcal^+,  
\label{eq : 4.a5}
\ee 
where the $P^A$ are two complex scalars. Note that
we are always allowed a spin transformation of the form 
$$
m^\alpha \mapsto \widetilde m^\alpha = e^{i \theta} \, m^\alpha, 
\qquad 
\om^\alpha \mapsto \widetilde \om^\alpha = e^{-i \theta} \, \om^\alpha, 
$$ 
where $\theta$ is any real-valued scalar field. 
This corresponds to the transformation  
$P^A \mapsto \widetilde P^A = e^{i \theta} \, P^A$, and 
thus $P^A$ and $\oP^A$ contain three real degrees of freedom, only. 
So, we define $P^A$ and $\oP^A$ by the conditions 
$$
P_A P^A = \oP_A \oP^A = 0, 
\qquad 
P_A \oP^A = \oP_AP^A = 1.  
$$
Then, defining 
$$
m_\alpha \, dx^\alpha = (P_B C^B) \, du - P_A \, dx^A \quad \mbox{ in } \Mcal^+
$$
completes the tetrad since $m_\alpha l^\alpha = m_\alpha n^\alpha = m_\alpha m^\alpha = 0$
and $m_\alpha\om^\alpha = -1$. 

Finally, we can obtain the general expression for the spacetime metric
on $\Mcal^+$, $g_{ab} = 2 \, l_{(a} \, n_{b)} - 2 \, m_{(a} \, \om_{b)}$, 
and summarize our conclusions, as follows.

\begin{lemma} 
\label{lemma1}
In the coordinates $(u,v,x^1,x^2)$ constructed above, the metric $g$ in both parts 
$\Mcal^\pm$ takes the general form 
$$
g = - 2 \, |P_A \, C^A|^2 \, du^2 + Q^{-1} \, du dv - h_{AB} \, C^A
\, du dx^B - h_{AB} \, dx^A dx^B, 
$$ 
where 
$$
h_{AB} = 2 \, P_{(A} \, \oP_{B)}
$$
is a real-valued metric on the two-surfaces $\Pcal_{u,v}$. 

When plane symmetry is imposed, one can choose $C^A =0$ identically, and
the metric only depends upon the function $Q$, and the symmetric $2$-tensor $h_{AB}$ 
$$
g = Q^{-1} \, du dv - h_{AB} \, dx^A dx^B; 
$$ 
moreover, all $\delta$- and $\odelta$-derivatives vanish. 
\end{lemma}

Note that the metric has six explicit {\sl degrees of freedom}. 
In this sense the coordinate choice is optimal. 
Also on $\Ncal_0$ we have $C^A=0$ by construction. 

Penrose's classification of the geometries allowed on $\Ncal_0$ can now be described as follows:
\begin{enumerate}

\item[$\bullet$] I-Geometry : The function  $Q$ to be discontinuous across $\Ncal_0$, while the coefficients
$C^A =0$ and $P^A$ are continuous. 

\item[$\bullet$] II-Geometry : The function $Q$ is continuous.

\item[$\bullet$] III-Geometry : The parameter $v$ is affine, so that $Q$ is constant on $\Ncal_0$
and without loss of generality one can assume that $Q \equiv 1$ on $\Ncal_0$. 

\end{enumerate}


\subsection{Jump relations for the NP connection scalars}

We now discuss solutions that may be singular across the hypersurface $\Ncal_0$, which has $v=0$ by 
definition. We will need to handle first- and second-order derivatives (in the variable $v$) 
of the coefficients $C^A$ and $P^A$ and, for instance,
our calculation will involve the jump $[ {C^A}_{,v} ]$ of the derivative of coefficient $C^A$ across the hypersurface 
$\Ncal_0$.

With Penrose's standard notation \cite{NP,Stewart91} and using the expression of the metric in Lemma~\ref{lemma1}, 
we obtain 
$$
\aligned
& \alpha = \half \oP_A \odelta P^A - \half \oP_A \delta \oP^A 
      - \tfrac{1}{4} \odelta \log Q - \tfrac{1}{4} \oP_A \Delta C^A,
\\      
& \beta = \half P_A \odelta P^A - \half P_A \delta \oP^A 
      - \tfrac{1}{4} \delta \log Q - \tfrac{1}{4} P_A \Delta C^A,
\\
& \gamma =  \tfrac{1}{4} \oP_A \Delta P^A - \tfrac{1}{4} P_A \Delta \oP^A, 
\\      
& \epsilon = \tfrac{1}{4} \oP_A D P^A + \tfrac{1}{4} P_A D \oP^A 
      - \tfrac{1}{4} \oP_A \delta C^A +\tfrac{1}{4} P_A \odelta C^A -
      \half D \log Q,
\endaligned
$$
$$
\aligned
& \kappa = 0, && \lambda = - \oP_A \Delta \oP^A, && 
\\
& \mu = - \half P_A \Delta \oP^A - \half \oP_A \Delta P^A, \qquad && \nu = 0,
\\
& \pi = - \half \odelta \log Q - \half \oP_A \Delta C^A, 
\endaligned
$$
and 
$$
\aligned
& \rho = \half \oP_A D P^A + \half P_A D \oP^A - \half \oP_A \delta C^A - \half P_A \odelta C^A, 
\\
& \sigma = P_A D P^A - P_A \delta C^A, 
\\
& \tau = \half \delta \log Q - \half P_A \Delta C^A,
\endaligned
$$
where the derivatives $D, \Delta, \delta, \odelta$ are understood in the sense of distributions. 

We have therefore established the following result. 

\begin{proposition}[Jump relations for Newman-Penrose connection scalars] 
\label{prop2} 
With the notation above, the NP connection scalars   
$\kappa$ and $\nu$ vanish identically while  $\rho, \sigma$, and $\epsilon$ are continuous across the
hypersurface $\Ncal_0$. The other NP connection scalars may be discontinuous across $\Ncal_0$, with jumps given by
$$
\aligned
& [\alpha] = -\tfrac{1}{4} \oP_{A} [C_v^A], 
&& [\beta] = [\oalpha], 
\\ 
& [\gamma] = \tfrac{1}{4} \, \oP_A \, [P_v^A] - \tfrac{1}{4} \, P_A \, [\oP_v^A],
&& [\lambda] = - \oP_A \, [\oP_v^A], 
\\
& [\mu] = - \half \, \oP_A \, [P_v^A] - \half \, P_A \, [\oP_v^A], 
\qquad && [\pi] = 2 [\alpha],
\\
& [\tau] = 2 [\oalpha]. 
\endaligned
$$
When plane symmetry is imposed one has also $\alpha= \beta=\pi=\tau=0$, so that the 
remaining jump conditions are
$$
\aligned 
& [\gamma] = \tfrac{1}{4} \, \oP_A \, [P_v^A] - \tfrac{1}{4} \, P_A \, [\oP_v^A],
\\
& [\lambda] = - \oP_A \, [\oP_v^A], 
\\
& [\mu] = - \half \, \oP_A \, [P_v^A] - \half \, P_A \, [\oP_v^A]. 
\endaligned
$$
\end{proposition}


\subsection{Jump relations for the NP curvature scalars}

One more piece of notation is now needed. 
If a scalar $f$ takes values $f^\pm$ on $\Ncal_0$ when approached from $\Mcal^\pm$, we set 
$$
\widetilde f = \half \, (f^+ + f^-),
$$
while we have already defined $[ f ] = f^+ - f^-$. We note the classical identity 
$$
[ f \, g ] = \widetilde f \, [ g ] + [ f ] \, \widetilde g.
$$ 
Recall that if $f$ contains a Dirac measure term, its coefficient is
denoted by $\otop f$.  

We now impose Einstein's field equations in the NP notation, as stated for instance in \cite{Stewart91} (p.~217).
By the equation (a) (in the notation therein) we find 
$$
\Phi_{00} = D \rho - \rho (\rho + \epsilon \oepsilon) - \sigma \osigma,
$$
whose right-hand side (according to Proposition~\ref{prop2}) is continuous across $\Ncal_0$, so that
$[\Phi_{00}] = 0$. 
Similarly, by equation (b) in \cite{Stewart91}  we find 
$$
\Psi_0 = D \sigma - (2 \rho + 3 \eps - \oepsilon) \sigma
$$
which implies $[\Psi_0 ] = 0$.  
The coefficient $\Phi_{22}$ is determined by equation (n): 
$$
\Phi_{22} = - \Delta \mu - \mu^2 - \lambda \olambda,
$$
which with Proposition~\ref{prop2} implies that  $\otop \Phi_{22} = - [\mu]$. 
Similarly, equation (j) yields 
$$
\Psi_4 = - \Delta \lambda - 2(\mu +  \gamma) \lambda,
$$
thus $\otop \Psi_4 = - [\lambda]$.

The coefficient $\Phi_{02}$ occurs in equations (g) and (p). 
Equation (p) involves the term $\Delta \sigma = \sigma_{,v}$, which
might have a non-zero jump across $\Ncal_0$ (which could also be computed
explicitly). However, by equation (g) we find $\otop \Phi_{20} = 0$ 
and 
$$
[\Phi_{20} ] = D [\lambda] -2 \odelta [\alpha] - (\rho - 3 \epsilon +
\oepsilon) [\lambda] - \osigma [\mu] - 2(2 \tildepi - \tilde\alpha + 
\tildeoverbeta) [\alpha].
$$
Next, $\Psi_1$ and $\Phi_{01}$ are involved in equations (c), (d),
(e), and (k).  
Equation (e) implies 
$$
\aligned
\otop\Psi_1 & = 0, 
\\
[\Psi_1] & = D [\oalpha] - 3 \sigma [\alpha] - (\rho - \oepsilon +
\epsilon) [\oalpha], 
\endaligned
$$  
while (d) gives 
$$
\aligned  
& \otop \Phi_{10} = 0, 
\\ 
& [\Phi_{10}] = D [\alpha] - (3 \rho -\epsilon + \oepsilon )  [\alpha] - 
\osigma [\oalpha]. 
\endaligned
$$
Equations (c) and (d) then provide no new information. 

The terms $\Psi_2$, $\Phi_{11}$, and $\Lambda$ occur in (f), (h), (l),
and (q).  
Unfortunately, (f) involves $\delta \epsilon$ while (q) involves
$\Delta \rho$, both of which are tedious to compute. 
We can however deduce  
$$
\otop \Psi_2 = \otop \Phi_{11} = \otop \Lambda = 0, 
$$ 
and the jump relations 
$$ 
\aligned 
\quad [\Psi_2]  + 2 [\Lambda] 
 =  
&D [\mu] - 2 \delta [\alpha] -  (\rho - \epsilon-\oepsilon) [\mu] 
\\
& - \sigma [\lambda]
- 2 (\tildeoverpi - \tildeoveralpha + \tildebeta) [\alpha] 
- 2 \tildepi [\oalpha], 
\endaligned
$$
and 
$$
\aligned
- [\Psi_2] + [\Lambda] + [\Phi_{11}] 
= & \delta [\alpha] - \odelta [\oalpha] 
 - \rho [\mu] + \sigma [\lambda]
\\
& - (\tildeoveralpha +3 \tildebeta) [\alpha] - 
(3\tildealpha + \tildeoverbeta) [\oalpha].  
\endaligned
$$ 

Finally, $\Psi_3$ and $\Phi_{12}$ occur in (i), (m), (o), and (r). 
Equation (r) provides us with 
$\otop \Psi_3 = - [\alpha]$, 
while (o) gives 
$\otop \Phi_{12} = - [\oalpha]$. 
Equation (i) is consistent with these results and gives no further information, but (m) reveals that 
$$
- [\Psi_3] + [\Phi_{21}] 
= \delta [\lambda] - \odelta [\mu] 
- (\tildealpha + \tildeoverbeta) [\mu] 
- (\tildeoveralpha - 3 \tildebeta) [\lambda]
- 2 \tildemu [\alpha] + 2 \tildelambda [\oalpha]. 
$$

We summarize our conclusions as follows. 

\begin{theorem} [Jump relations for Newman-Penrose curvature scalars]
Assuming that the hypersurface $\Ncal_0$ has type II-geometry, then the
Dirac measure terms on it satisfy the following conditions: 
$$
\aligned
& \quad  \otop \Psi_4 = - [\lambda], \quad \otop \Psi_3 = - [\alpha], 
\quad \otop \Phi_{22} = - [\mu], \quad  \otop \Phi_{12} = - [\oalpha], 
\\
& \otop \Phi_{00} =
\otop \Phi_{10} = \otop \Phi_{20} = \otop \Phi_{11} = \otop \Lambda = 0,
\\
&  \otop \Psi_2 = \otop \Psi_1 = \otop\Psi_{0}= 0.  
\endaligned
$$
If one strengthens the assumption to that of type III-geometry, i.e.,
$[\alpha]=0$, then in addition $ \otop \Psi_3=\otop \Phi_{12} = 0$. When plane-symmetry is imposed the latter condition
are automatically satisfied and the geometry is always of type III. 
If one considers ``pure gravitational radiation'' which, following Penrose, 
requires no Dirac measure terms in the trace-free Ricci tensor, then one
has $[\mu]=0$ as well.
\end{theorem}
 
 These results can be interpreted in terms of physical quantities, as follows. 
For instance, a
concentration (Dirac mass structure) in the
energy density of the fluid requires a jump in the Ricci curvature, their relative strength being proportional. This statement, in particular, is valid as one approaches 
 the future boundary of the maximal development (considered in the previous section) 
and provides a corresponding relation between the curvature and matter content of the spacetime.


\section*{Acknowledgments}

The authors are grateful to the referees for a detailed reading of this paper and many  constructive suggestions. The first author (PLF) was partially supported by  
the Agence Nationale de la Recherche (ANR) through the grant 06-2-134423 entitled
``Mathematical Methods in General Relativity'', the Centre National de la Recherche Scientifique (CNRS), and the Erwin Schr\"odinger Institute (Vienna).



\begin{thebibliography}{3} 

\bibitem{AS65} Abramowitz M. and Stegun I.A., 
{\em Handbook of mathematical functions,} Dover, New York, 1965.

\bibitem{AB} Aichelburg P.C. and Balasin H.,
Generalized symmetries of impulsive gravitational waves. Geometry and physics,
Class. Quantum Grav. 14 (1997), A31--A41. 

\bibitem{AS} Aichelburg P.C. and Sexl R.U., 
On the gravitational field of a massless particle,
J. Gen. Relat. Grav. 2 (1971), 303--312. 

\bibitem{ALF} Amorim P. and LeFloch P.G., 
Sharp estimates for periodic solutions to the Euler--Poisson--Darboux equation, 
Port. Math. 65 (2008), 387--429. 

\bibitem{BLSS} Barnes A.P., LeFloch P.G., Schmidt B.G., and Stewart J.M., 
The Glimm scheme for perfect fluids on plane-symmetric Gowdy spacetimes,
Class. Quantum Grav. 21 (2004), 5043--5074.  

\bibitem{CX} Chandrasekhar S. and Xanthopoulos B.C., 
On the collision of impulsive gravitational waves when coupled with
fluid motions,  Proc. Royal Soc. Lond. A  402 (1985),  37--65.  

\bibitem{Choquet-book} Choquet-Bruhat Y.,
{\em General relativity and the Einstein equations,}
Oxford Math. Monographs, Oxford Univ. Press, 2009.  
  
\bibitem{Christo00} Christodoulou D.,  
The structure and uniqueness of generalized solutions of the spherically symmetric Einstein-scalar equations,
Comm. Math. Phys. 109 (1987), 591--611. 
  
\bibitem{Christo0} Christodoulou D.,
Global existence of generalized solutions of the spherically symmetric Einstein-scalar equations in the large,
Comm. Math. Phys. 106 (1986), 587--621. 
  
\bibitem{Christo1} Christodoulou D., 
Bounded variation solutions of the spherically symmetric
Einstein-scalar field equations,  Comm. Pure Appl. Math. 46 (1992), 1131--1220.

\bibitem{Christo2} Christodoulou D., 
Self-gravitating relativistic fluids: a two-phase model, 
Arch. Ratio. Mech. Anal. 130 (1995), 343--400.  

\bibitem{Christodoulou5} Christodoulou D., 
{\em The formation of black holes in general relativity,}
EMIS, 2009. 

\bibitem{CH62} Courant R. and Hilbert D., 
{\em Methods of mathematical physics,} Vol. 2, 1962, Wiley-Interscience.

\bibitem{Darboux} Darboux G., 
{\em Le\c{c}ons sur la th\'eorie g\'en\'erale des surfaces}, Vol. II, 1899, Gauthier, Paris.

\bibitem{Feinstein} Feinstein A., Kunze K.E., and V\'azquez-Mozo M.A.,
Initial conditions and the structure of the singularity in pre-big-bang cosmology,
Class. Quantum Grav. 17 (2000), 3599--3616.

\bibitem{Ferrari} Ferrari V., Pendenza P., and Veneziano G.,
Beam-like gravitational waves and their geodesics,
Gen. Relat. Grav. 20 (1988), 1185--1191. 

\bibitem{Friedrich} Friedrich H.,
On the regular and the asymptotic characteristic initial value problem
for Einstein's vacuum field equations,
Proc. Roy. Soc. London Ser. A 375 (1981), 169--184. 

\bibitem{Griffiths} Griffiths J.B., 
{\em Colliding plane waves in general relativity,}
Oxford Univ. Press, 1991. 

\bibitem{GSR} Griffiths J.B. and Santano-Roco M., 
The characteristic initial value problem for colliding plane waves:
the linear case,  Class. Quantum Grav. 19 (2002), 4273--4286.  

\bibitem{HE} Hauser I and Ernst F.J., 
Initial value problem for colliding plane waves I,
J. Math. Phys. 30 (1989), 872--887.
  
\bibitem{KhanPenrose} Khan K. and Penrose R., 
Scattering of two impulsive gravitational plane waves, Nature 229
(1971), 185--186.  

\bibitem{KS} Kunzinger M. and Steinbauer R.,
A rigorous solution concept for geodesic and geodesic deviation equations in impulsive gravitational waves,
J. Math. Phys. 40 (1999), 1479--1489. 
 
\bibitem{LM} LeFloch P.G. and Mardare C., 
Definition and weak stability of spacetimes with distributional curvature, 
Port. Math. 64 (2007), 535--573.

\bibitem{LR} LeFloch P.G. and Rendall A.D., 
A global foliation of Einstein-Euler spacetimes with Gowdy-symmetry on $T^3$,  
Arch. Rational Mech. Anal. (2011). See also ArXiv:1004.0427.

\bibitem{LeFlochSmulevici} LeFloch P.G. and Smulevici J., 
 in preparation. 

\bibitem{LS} LeFloch P.G. and Stewart J.M., 
Shock waves and gravitational waves in matter spacetimes with Gowdy symmetry, 
Portugal. Math. 62 (2005), 349--370. 

\bibitem{Moncrief} Moncrief V., 
Global properties of Gowdy spacetimes with T3 topology, Ann. Phys. 132 (1981), 87--107. 

\bibitem{NP} Newman E.T. and Penrose R., 
An approach to gravitational radiation by a method of spin coefficients, 
J. Math. Phys. 3 (1962), 566--578. 

\bibitem{Olver} Olver F.W.J., 
{\em Asymptotics and special functions,} 
  Academic Press, New York, 1974.

\bibitem{Penrose0} Penrose R., 
A remarkable property of plane wave in general relativity, Rev. Modern
Phys. 37 (1965), 215--220. 

\bibitem{Penrose} Penrose R., 
The geometry of impulsive gravitational waves,
in: ``General Relativity, Papers in honour of J.L. Synge'', 
ed. L. O'Raifeartaigh, 1972, Clarendon Press, Oxford, pp. 101--115. 

\bibitem{Rendall} Rendall A.D., 
Reduction of the characteristic initial value problem to the Cauchy
problem and its applications to the Einstein equations, 
Proc. R. Soc. Lond. A 427 (1990), 221--239. 
 
\bibitem{Stellmacher} Stellmacher K., 
Ausbreitungsgesetze f\"ur charakteristische Singularit\"aten der
Gravitationsgleichungen, Math. Annalen 115 (1938), 740--783.

\bibitem{Stewart91} Stewart J.M.,  
{\sl Advanced general relativity}, Cambridge Univ. Press, 1991.

\bibitem{Stewart09} Stewart J.M., 
The Euler-Poisson-Darboux equation for relativists, 
Gen. Relat. Grav. 41 (2009), 2045--2071. 

\bibitem{SF} Stewart J.M. and Friedrich H., 
Numerical relativity. The characteristic initial value problem, 
Proc. R. Soc. Lond. A 384 (1982), 427--454. 

\bibitem{Szekeres} Szekeres P., 
Colliding plane gravitational waves, J. Math. Phys. 13 (1972), 286--294. 

\bibitem{TT} Tabensky R. and Taub A.H., 
Plane symmetric self-gravitating fluids with pressure equal to energy density, 
Comm. Math. Phys. 29 (1973), 61--77. 

\bibitem{Taub} Taub A.H., 
General relativistic shock waves in fluids for which pressure equals
energy density,  Comm. Math. Phys. 29 (1973), 79--88. 

\bibitem{VW} Veneziano G. and Wosiek J.,
Exploring an $S$-matrix for gravitational collapse,
J. High Energy Phys. 9 (2008), 023--038.

\bibitem{Yurtsever1} Yurtsever U., 
Structure of the singularities produced by colliding plane waves, 
Phys. Rev. D 38 (1988),  1706--1730. 

\bibitem{Yurtsever2} Yurtsever U., 
Singularities and horizons in the collision of gravitational waves, 
Phys. Rev. D 40 (1989), 329--359. 

\bibitem{Wald} Wald R.M., 
{\sl General relativity,} University of Chicago Press, 1984. 
 
\end{thebibliography}
\end{document}